\PassOptionsToPackage{table,xcdraw}{xcolor}
\documentclass[acmlarge,anonymous=false]{acmart}
\usepackage{booktabs}
\usepackage{multirow}  
\usepackage{subfigure}
\usepackage{graphicx}
\usepackage{lipsum}
\usepackage{textcomp}
\usepackage{xcolor}
\AtBeginDocument{%
  \providecommand\BibTeX{{%
    \normalfont B\kern-0.5em{\scshape i\kern-0.25em b}\kern-0.8em\TeX}}}

\setcopyright{acmcopyright}
\copyrightyear{2020}
\acmYear{2020}
\acmDOI{00.0000/00.00000}

\acmJournal{IMWUT}
\acmVolume{4}
\acmNumber{3}
\acmArticle{0}
\acmMonth{0}

\begin{document}

\title{n-Gage: Predicting in-class Emotional, Behavioural and Cognitive Engagement in the Wild}


\author{Nan Gao}
\email{nan.gao@rmit.edu.au}
\affiliation{%
  \institution{RMIT University}
  \city{Melbourne}
  \country{Australia}
  \postcode{3000}
}
\author{Wei Shao}
\email{wei.shao@rmit.edu.au}
\affiliation{%
  \institution{RMIT University}
  \city{Melbourne}
  \country{Australia}
  \postcode{3000}
}
\author{Mohammad Saiedur Rahaman}
\email{saiedur.rahaman@rmit.edu.au}
\affiliation{%
  \institution{RMIT University}
  \city{Melbourne}
  \country{Australia}
  \postcode{3000}
}

\author{Flora D. Salim}
\email{flora.salim@rmit.edu.au}
\affiliation{%
  \institution{RMIT University}
  \city{Melbourne}
  \country{Australia}
  \postcode{3000}
}

\begin{abstract}
  
  The study of student engagement has attracted growing interests to address problems such as low academic performance, disaffection, and high dropout rates. Existing approaches to measuring student engagement typically rely on survey-based instruments. While effective, those approaches are time-consuming and labour-intensive. Meanwhile, both the response rate and quality of the survey are usually poor. As an alternative, in this paper, we investigate whether we can infer and predict engagement at multiple dimensions, just using sensors. We hypothesize that multidimensional student engagement level can be translated into physiological responses and activity changes during the class, and also be affected by the environmental changes. Therefore, we aim to explore the following questions: \textit{Can we measure the multiple dimensions of high school student's learning engagement including emotional, behavioural and cognitive engagement with sensing data in the wild? Can we derive the activity, physiological, and environmental factors contributing to the different dimensions of student learning engagement? If yes, which sensors are the most useful in differentiating each dimension of the engagement?}
  Then, we conduct an in-situ study in a high school from 23 students and 6 teachers in 144 classes over 11 courses for 4 weeks. We present the \textit{n-Gage}, a student engagement sensing system using a combination of sensors from wearables and environments to automatically detect student in-class multidimensional learning engagement. Extensive experiment results show that \textit{n-Gage} can accurately predict multidimensional student engagement in real-world scenarios with an average mean absolute error (MAE) of 0.788 and root mean square error (RMSE) of 0.975 using all the sensors. We also show a set of interesting findings of how different factors (e.g., combinations of sensors, school subjects, $CO_2$ level) affect each dimension of the student learning engagement.

\end{abstract}

\ccsdesc[500]{Human-centered Computing~Ubiquitous and mobile computing}
\ccsdesc[300]{Applied computing~Education}

\keywords{Engagement; Students; Wearable; Electrodermal Activity}

\maketitle

\section{Introduction}
 In education, \textit{student engagement} refers to the degree of attention, interest, curiosity, and involvement in the learning environment \cite{groccia2018student}. The study of student engagement has attracted growing interests as a way to address the problems of low academic achievement, high levels of student boredom, disaffection, and high dropout rates in urban areas \cite{fredricks2004school,fredricks2012measurement}. Previous research showed that student engagement declines as students progress from elementary to middle school, reaching its lowest levels in high school \cite{martin2016student,national2003engaging,marks2000student}. Marks et al. \cite{marks2000student} estimated that as many as 40-60\% of high school students are disengaged (e.g., uninvolved, no interests, and not attentive). The consequences of disengagement for high school students are severe. They are less likely to graduate from high school and face limited employment prospects, increasing risks for poverty, poorer health, and involvement in the criminal justice system \cite{national2003engaging}. Given the negative impact of disengagement, more and more researchers, educators, and policymakers are interested in obtaining data on student engagement and disengagement for needs assessment, diagnosis, and preventive measures \cite{martin2016student}. 
 
Generally, student engagement is defined as a meta-construct that includes three dimensions \cite{fredricks2004school,fredricks2012measurement}: 
(1) \textit{behavioural engagement} focuses on participation and involvement in academic, social, and co-curricular activities. Some researchers define behavioural engagement with regards to positive conduct, e.g., following the rules, and the absence of disruptive behaviour such as skipping school \cite{fredricks2012measurement, finn1995disruptive, finn1997academic};
(2) \textit{emotional engagement} focuses on the extent of positive and negative reactions to teachers, classmates, academics, and school, which includes a sense of belonging or connectedness to the school \cite{fredricks2012measurement, finn1989withdrawing};
(3) \textit{cognitive engagement} draws on the idea of investment in learning. It incorporates thoughtfulness and willingness to put effort to comprehend complex ideas and master difficult skills \cite{fredricks2004school, fredricks2012measurement,corno1983role}. One of the widely used method for measuring student engagement is self-report survey, e.g., \textit{Motivated Strategies for Learning Questionnaire} (MSLQ) \cite{pintrich1990motivational}, \textit{School Engagement Measure} (SEM) \cite{moore2005conceptualizing}, and \textit{Engagement vs. Disaffection with Learning} (EvsD) \cite{skinner2009motivational}. Though generally reliable, the survey is  time-consuming and may become a burden for participants if they need to complete 
it for each class. 

Therefore, we want to investigate whether we can infer and predict multidimensional student engagement just using sensors. In particular, we conduct the research around the following hypothesis: the multidimensional student engagement level can be translated into physiological responses and activity movements during the class, and can also be affected by environmental changes. In previous studies, various physiological data, (e.g., electrodermal activity (EDA), heart rate variability (HRV), accelerometer (ACC), skin temperature (ST)) and environmental data have been explored to assess the emotional arousal and engagement level in different scenarios. 
For instance, EDA is usually considered as a good indicator of psychological or physiological arousal (e.g., emotional and cognitive states \cite{critchley2002electrodermal, boucsein2012electrodermal}) , which has been increasingly explored in affective computing, such as the detection of emotion \cite{bakker2011s, canento2011multimodal}, depression \cite{sarchiapone2018association} , and engagement \cite{di2018unobtrusive, hernandez2014using, latulipe2011love}. Recently, Pflanzer et al. \cite{pflanzer2013galvanic} stated that EDA monitoring should be combined with the heart rate because they are both autonomically dependent variables. Heart rate has been used for student engagement prediction \cite{monkaresi2016automated} and the correlation of heart rate and cognitive/emotional engagement has been found in \cite{fuller2018development}. As the most commonly used sensor in IoT devices, accelerometer is proven to be powerful for quantifying human behavioural patterns \cite{wang2014studentlife, gao2019predicting}. It has been used for demonstrating how synchronized movement of people can enhance group affiliation \cite{von2018choreography} and sensing engagement using interpersonal movement synchrony \cite{ward2018sensing}. 



In this paper, our research questions are as follows: 1. \textit{Can we measure the multiple dimensions of high school student's learning engagement including emotional, behavioural and cognitive engagement with sensing data in the wild?} 2. \textit{Can we derive the activity, physiological, and environmental factors contributing to the different dimensions of student learning engagement? If yes, which sensors are the most useful in differentiating each dimension of the engagement?} To 
answer the above questions and enable automated engagement detection, we present a new classroom sensing system \textit{n-Gage} to assess the behavioural, emotional and cognitive engagement levels of high school students. The system utilizes sensing data from two sources: (1) wearable devices capturing physiological and physical signals (e.g., EDA, HRV, ACC, ST); (2) indoor weather stations capturing environmental changes (e.g., temperature, CO$_2$, sound). The study has been approved by the Human Research Ethics Committee at our University and the high school where it is conducted, and all the procedures follow the ethical codes. The main contributions of this paper include:

\begin{itemize}
    \item We collect a dataset of 23 high school students and 6 teachers in 144 classes over 11 courses for 4 weeks. Weather stations are installed in 3 classrooms, and student participants are asked to wear the E4 wristbands and complete online survey 3 times a day to report their behavioural, emotional and cognitive engagement level during the classes. To the best of our knowledge, this is the most diverse and largest dataset collected in the wild to measure student engagement using sensors. 
    
    \item We build \textit{n-Gage}, a classroom sensing system to automatically measure the multidimensional engagement (behavioural, emotional and cognitive engagement) of high school students during the classes. In particular, we combine physiological signals, physical activities, and indoor environmental data to estimate the changes in student engagement levels. To the best of our knowledge, this is the first system to detect student engagement from multiple sensors in the wild.
    
    \item We extract new features to represent the physiological and physical synchrony between students which proved to be useful for the student engagement prediction. For the first time, we extract features from skin temperature and indoor environment for effective engagement estimation.
    
    \item We conduct comprehensive experiments to predict multidimensional student engagement scores with \textit{LightGBM} regressors. The experiment results show that \textit{n-Gage} reaches a high accuracy (0.563 MAE and 0.715 RMSE score) for student engagement prediction. We also derive different factors and explore the most useful sensors in differentiating each dimension of the learning engagement.
    \item We show a set of interesting insights on how different factors affect student engagement. For example, the CO$_2$ level in the classroom has a negative impact on students' cognitive engagement, which highlights the need to ventilate the classroom timely to improve student engagement.
    
    
\end{itemize}

The remainder of the paper is as follows. Section ~\ref{sec:related work} introduces related works of traditional methods for measuring engagement, and the recent progress of engagement prediction with sensing technology. Section ~\ref{sec: data collection} describes the data collection procedures, including participant recruitment, equipments for data collection, and the self-report instrument. Then we introduce data pre-processing techniques in Section \ref{sec: preprocessing}. In Section ~\ref{sec: data analysis}, we extract various features from physiological signals and environmental changes. Section ~\ref{sec: prediction pipeline} introduces the prediction pipeline and Section ~\ref{sec:results} shows experiment results and in-depth discussion about engagement prediction. Section ~\ref{sec:implications and limitations} lists the implications and limitations of our work. Finally, we summarize this research in Section ~\ref{sec:conclusion} and indicate the potential direction of future work.

\section{Related Work}
\label{sec:related work}

\subsection{Traditional Methods for Measuring Engagement}
In the education area, there are various methods to study student engagement. 
(1) \textit{Student Self-report} is the most common method to assess student engagement as it is easy to execute in classroom settings. Students are provided with items reflecting different dimensions of engagement and then select the response that best describes them \cite{fredricks2012measurement}. However, the self-report survey is labour and time-consuming, and students may not be willing to answer too many questions honestly at a time, leading to low-quality responses \cite{appleton2006measuring}.
(2) \textit{Experience Sampling} \cite{shernoff2014student} allows researchers to collect responses at the moment, which reduces the problems of recall-failure and social-desirability bias happened in the self-report surveys. However, it requires a huge time investment from students, and the quality of responses largely relies on the students' willingness and ability to answer \cite{fredricks2012measurement}.
(3) \textit{Teacher Ratings of Students} \cite{fredricks2012measurement} can be useful for young students with difficulty in completing self-report surveys.  Behaviour can be observed directly from teachers, but emotion engagement is difficult to be observed as students may learn to mask their emotions \cite{fredricks2012measurement,skinner2008engagement}.
(4) \textit{Interviews} can provide a detailed description of the student's performance during the learning process. However, the quality of responses depends on the expert knowledge from the interviewers. 
(5) \textit{Observations} \cite{fredricks2012measurement} on the individual student or whole students in the classroom have been developed to assess engagement, which can be time-consuming for the administration and all kinds of academic settings need to be considered to get an accurate picture of student behaviour. The reliability of the observations can be doubtful as they only provide limited information about students.

All traditional methods for engagement measurement have strengths and limitations in different situations. Overall, traditional methods are usually time-consuming, and the quality of answers largely depends on the students, teachers, or executor. Recently, with the development of wearables and IoT sensors, some initial progress has been explored to measure student engagement with physiological signals which is more subjective and obtrusive to students.   

\subsection{Engagement Prediction with Sensing Technology}

\begin{table}[]
\caption{Related works for engagement prediction with sensing data}
\label{tab:related works}

\begin{tabular}{@{}llll@{}}
\toprule 

\textbf{Prediction}                                                                                              & \textbf{Data source}                                                            & \textbf{Participants}                                                                           & \textbf{Data Sessions} \\ [0.3em]
\rowcolor[HTML]{EFEFEF} 
Audience Engagement \cite{ward2018sensing}                                                                                    & ACC data                                                                        & \begin{tabular}[c]{@{}l@{}}10 children audience\\  in art performance\end{tabular}              & not stated             \\

Social Engagement \cite{hernandez2014using}                                                                                      & EDA data (wristband)                                                            & \begin{tabular}[c]{@{}l@{}}Children during social\\  interactions\end{tabular}                  & 51 sessions            \\
\rowcolor[HTML]{EFEFEF} 
Game Engagement \cite{huynh2018engagemon}                                                                                       & EDA, PPG data                                                                   & \begin{tabular}[c]{@{}l@{}}10 players in 6 mobile \\ games in natual settings\end{tabular}      & not stated             \\

Audience Engagement \cite{gashi2019using}                                                                                    & EDA, PPG data                                                                   & \begin{tabular}[c]{@{}l@{}}10 attendess and 19 \\ presentors in presentations\end{tabular}   & 40 sessions            \\
\rowcolor[HTML]{EFEFEF} 
Student Engagement \cite{ahuja2019edusense}                                                                                      & Video, audio data                                                               & \begin{tabular}[c]{@{}l@{}}25 university students\\  in 5 classrooms\end{tabular}               & not stated             \\

Student Engagement \cite{monkaresi2016automated}                                                                                   & Video, heart rate data                                                          & \begin{tabular}[c]{@{}l@{}}23 university students \\ in laboratory settings\end{tabular}        & not stated             \\
\rowcolor[HTML]{EFEFEF} 
Student Engagement \cite{mcneal2014measuring}                                                                                     & EDA data (hand sensor)                                                          & \begin{tabular}[c]{@{}l@{}}17 undergraduate students\\  in climate science classes\end{tabular} & not stated             \\

Student Engagement \cite{wang2015physiological}                                                                                     & EDA data (hand sensor)                                                          & \begin{tabular}[c]{@{}l@{}}17 university students in \\ learning environments\end{tabular}      & not stated             \\
\rowcolor[HTML]{EFEFEF} 
Emotional Engagement \cite{di2018unobtrusive}                                                                                   & EDA data (wristband)                                                            & \begin{tabular}[c]{@{}l@{}}27 university students in \\ 41 lectures over 3 weeks\end{tabular}   & 197 sessions           \\

\begin{tabular}[c]{@{}l@{}}Multidimensional Student \\Engagement (this work)\end{tabular} & \begin{tabular}[c]{@{}l@{}}EDA, PPG, ST, ACC, \\ CO2,  Noise, etc.\end{tabular} & \begin{tabular}[c]{@{}l@{}}23 high school students in\\  98 classes over 4 weeks\end{tabular}   & 331 sesssions          \\ \bottomrule
\end{tabular}
\end{table}
Sensing technologies are becoming prevalent to assess people's mental characteristics (e.g., engagement \cite{di2018unobtrusive,hernandez2014using,ward2018sensing,hutt2019automated, huynh2018engagemon}, mood \cite{morshed2019prediction,wang2014studentlife}, stress \cite{bakker2011s, wang2014studentlife}, personality \cite{gao2019predicting}) and have provided an attractive alternative to traditional self-report surveys. Wang et al. \cite{wang2014studentlife} gathered students' mental health data such as mood and stress from self-report surveys in Dartmouth college. They also recorded students' activity data from passive sensors and found a significant correlation between the sensor data and mental health. Morshed et al. \cite{morshed2019prediction} predicted mood instability only using sensed data from mobile phones and wearable sensors for individuals in situated communities. Wang et al. \cite{wang2018sensing} predicted human personality traits from passive sensing data from mobile phones using within-person variability features such as regularity index of physical activity, the circadian rhythm of location. 

Physiological sensors and accelerometers have been explored to assess human's engagement (see Table ~\ref{tab:related works}), such as assessing audience engagement during the art performance, social engagement for children during the interaction with adults \cite{hernandez2014using}, emotional engagement for university students during lectures \cite{di2018unobtrusive}. 

Ahuja et al. \cite{ahuja2019edusense} built a classroom sensing system with commodity cameras. Students' and instructor's video and audio were captured for body segmentation and speech detection. Then, the students' engagement levels were analyzed based on their facial expressions and gestures. However, as reported from authors, this system would bring privacy concerns \cite{kjaergaard2020current} when capturing audio and video data. Similarly, Hutt et al. \cite{hutt2019automated} used commercial off-the-shelf eye-trackers to detect mind wandering for high school students and Monkaresi et al. \cite{monkaresi2016automated} used heart rate and video-based estimation of facial expressions to predict the engagement of 23 university students during a structured writing activity in laboratory settings. 

Only a few studies investigate student engagement in real-world settings \cite{mcneal2014measuring,wang2015physiological, fuller2018development,di2018unobtrusive}. Mcneal et al. \cite{mcneal2014measuring} used EDA hand-sensors to measure the engagement from 17 undergraduate students in climate science classes during a semester. They explored different teaching approaches on a subset of students and reported the statistical mean value of EDA traces. Contrast to their study, we collect a far more heterogeneous data set and novel features were proposed based on different physiological indices. Wang et al. \cite{wang2015physiological} studied 17 university students' engagement in the distributed learning environment with EDA hand sensors, and found that EDA measurements were aligned with surveys. Different to our research, they only used a very simple question 'how much did you enjoy during the lecture' as the ground truth of students' engagement. 

In recent years, researchers have started to explore different dimensions of engagement using physiological signals. Lascio et al. \cite{di2018unobtrusive} predicted university students' emotional engagement from EDA sensors in lectures during 3-week data collection. While in our data collection, we build an in-class multidimensional (behavioural, emotional, cognitive) engagement sensing system including physiological responses (i.e., EDA, HRV, ST), physical movements (ACC) and indoor environmental sensors (i.e., CO$_2$, temperature, humidity, sound) for high school students. Furthermore, high school classes are very different from lectures at university in \cite{di2018unobtrusive} (e.g., degree of freedom to choose courses, ability to schedule classes flexibly, requirements of class attendance, consistency of subjects between different schools), which may lead to very different multi-engagement distribution in high school classes. Another similar research, but for a different application, was proposed by Huynh et al. \cite{huynh2018engagemon} who measured the engagement level of game players with multiple sensors. Though they agreed that user engagement includes three dimensions, they did not differentiate each dimension when predicting the engagement during the game. Nevertheless, in our study, we derive the different factors and most useful sensors contributing to the different dimensions of student learning engagement.


In summary, different from previous efforts, our work has several advantages: (1) we use far more heterogeneous data for engagement prediction (others only use EDA or heart rate data except \cite{huynh2018engagemon}); (2) we propose and extract more meaningful features from physiological signals while \cite{wang2015physiological,mcneal2014measuring,fuller2018development} only use the simple average value of data; (3) to the best of knowledge, we are the first to predict the engagement for all three dimensions based on education research while previous studies either measure the simple general engagement, or a single dimension of engagement \cite{di2018unobtrusive}), and derive the most useful sensors in differentiating each dimension of engagement; (5) we adopt real-world classroom settings and take the influence of environmental changes into account. 

\section{Data Collection}
\label{sec: data collection}
We conducted a field study in a private high school for 4 weeks in 2019. The data collection has been approved by the Human Research Ethics Committee at our University. We will then provide the details about participants, equipments to collect the data, and data collection procedures. 

\subsection{Participants}

\begin{figure}[b]
	\centering
	\subfigure[Empatica E4 wristbands \label{fig: 4band}]{\includegraphics[width=0.3\textwidth]{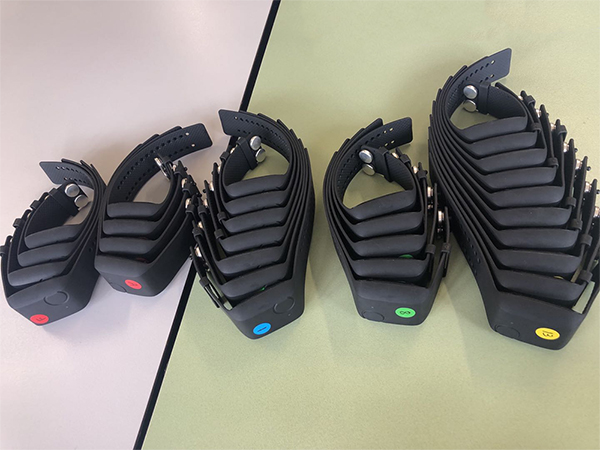}
	\label{fig: sub e4band}}
	\hspace{0.2cm}
	\subfigure[Netatmo indoor weather station]{\includegraphics[width=0.3\textwidth]{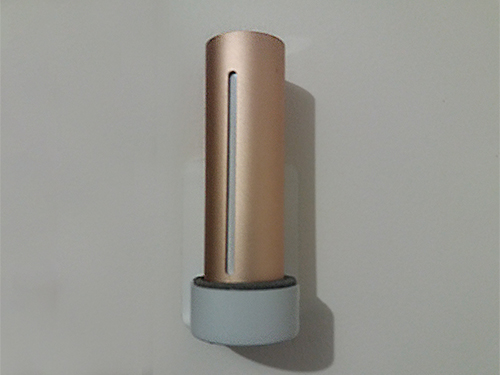}
    \label{fig: sub netamo}}
    \hspace{0.2cm}
    \subfigure[Classroom for Year 10 students]{\includegraphics[width=0.3\textwidth]{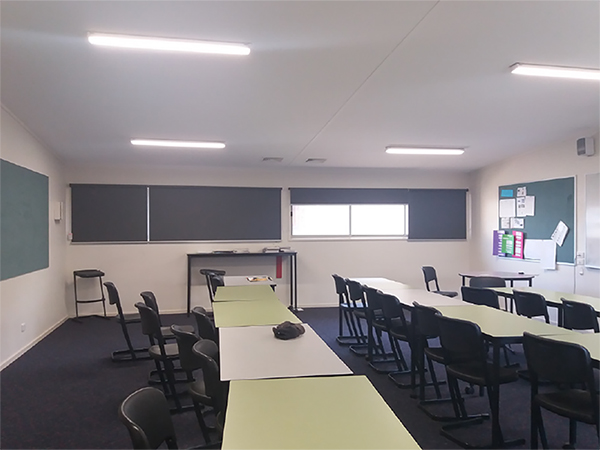}   
    \label{fig: sub room}}
    \caption{Devices and environments for collecting wearable and indoor data}
    \label{figb: netamo and e4band}
\end{figure}

We recruited 23 students (13 females and 10 males, 15-17 years old) and 6 teachers (4 females and 2 males, 33-62 years old) in Year 10 (see Table ~\ref{tab:info}). First, we gave an introduction to all Year 10 students and teachers, and distributed consent forms to them. Then, students or teachers who volunteered to participate returned the signed consent form from themselves and their (students') guardians. Once they signed forms, they were asked to complete the online background survey. Background information was collected at once including age, gender, how their engagement is affected by their thermal feelings, form class group, math class group, and language class group. There are 3 form classes in Year 10 and students are in form groups when having most classes (i.e., English, Global Politics, Science, Physical Education, Health/Sport). For Mathematics class, students are divided into 3 study groups and students in each group take classes in individual classrooms. For Language class, students have 4 different groups such as Japanese, French, Cultural Sustainability, and each group has classes in different classrooms too. Hence, the background information of different study groups helps us align students in different classrooms (see Figure ~\ref{fig: sub room}) at different times. We also recruited 3 Math teachers, 1 English teacher, 1 Japanese teacher and 1 Science teacher. Table \ref{tab:classinfo} shows the details about room allocation for participants.

\begin{table}
\centering
\begin{minipage}[]{6cm}
\caption{Room allocation for different class groups. Most classes belong to form group.}
\label{tab:classinfo}
\begin{tabular}{@{}llll@{}}
\toprule
\multirow{2}{*}{\textbf{Room}} & \multicolumn{3}{c}{\textbf{Number of Students}}   \\ \cmidrule(l){2-4} 
                      & \textit{Form} & \textit{Math} & \textit{Language  } \\ \cmidrule(r){1-4}
Room 1                & 10         & 7          & 13             \\
Room 2                & 6          & 9          & 2              \\
Room 3                & 7          & 7          & 5              \\
Room 4                & NaN          & NaN          & 3              \\ \bottomrule
\end{tabular}
\end{minipage}
\hspace{1.8cm}
\begin{minipage}[]{6cm}
\caption{Basic information for student and teacher participants.}
\label{tab:info}
\begin{tabular}{@{}lll@{}}
\toprule
\textbf{Category} & \textbf{Students} & \textbf{Teachers} \\ \midrule
Total Number      & 23                & 6                 \\
Female            & 13                & 4                 \\
Male              & 10                & 2                 \\
Age               & 15-17 y.o.           & 33-62 y.o.             \\ \bottomrule
\end{tabular}

\end{minipage}
\end{table}

\subsection{Collected Data}

\subsubsection{Physiological and Activity Data} 
During the school time, we asked participants to wear \textit{Empatica E4} \footnote{Empatica E4 wristband: \url{https://www.empatica.com/en-int/research/e4/}} wristbands as shown in Figure ~\ref{fig: sub e4band}, first proposed in \cite{garbarino2014empatica}. E4 wristband is a watch-like device with multiple sensors: electrodermal activity (EDA) sensor, photoplethysmography (PPG) sensor, 3-axis accelerometer (ACC), and optical thermometer. EDA depicts constantly fluctuating changes in skin electrical properties at 4 Hz. When the level of sweat increases, the conductivity of skin increases. PPG sensor measures the blood volume pulse (BVP) at 64 Hz, from which the inter-beat interval (IBI) and heart rate variability (HRV) can be derived. ACC records 3-axis acceleration in the range of [-2g, 2g] at 32Hz and captures motion-based activity. The optical thermometer reads peripheral skin temperature (ST) at 4 Hz. In the recording mode, E4 wristband can store 60 hours of data in the memory, and the battery can last for more than 32 hours. It is light-weight, comfortable and water-proof, thus especially suitable for continuous and unobtrusive monitoring of participants in our study. 

\subsubsection{Indoor Environmental Data}
\label{sec: indoor environmental data}
\begin{figure}
    \centering
    \subfigure[Temperature]{\includegraphics[width=0.48\textwidth]{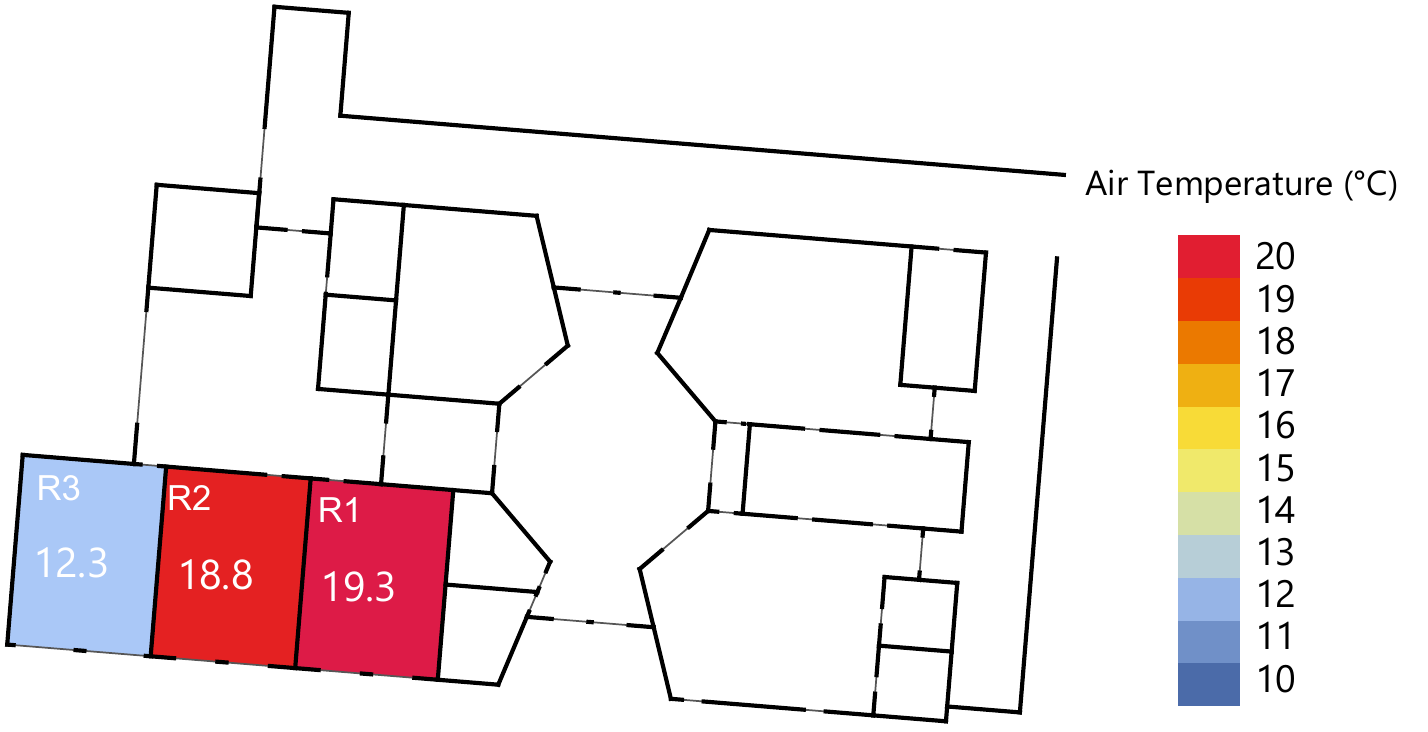}}
    \subfigure[CO$_2$ level]{\includegraphics[width=0.48\textwidth]{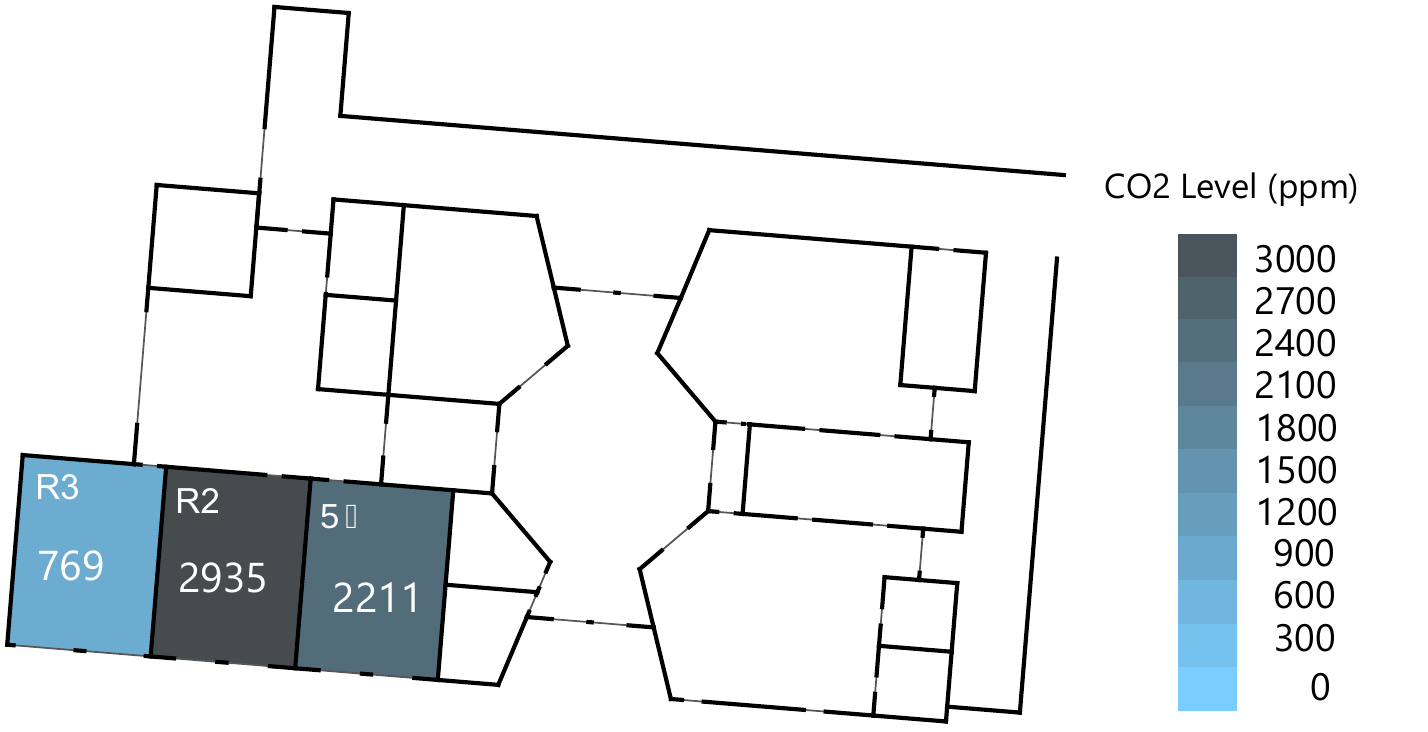}}
    \caption{Temperature and CO$_2$ data in R1, R2, R3 (room 1, room 2, room 3) at 11:00 am on 11 Sep 2019. Room 4 is not shown here as it is in another building.}
    \label{figb:co2 and temp}
\end{figure}
We collected indoor environmental data from the Netatmo Healthy Home Coach \footnote{Netatmo Healthy Home Coach: \url{https://www.netatmo.com/en-eu/aircare/homecoach}} - a smart indoor weather station - installed in the classrooms as shown in Figure ~\ref{fig: sub netamo} and Figure ~\ref{fig: sub room}. The Netatmo station can collect indoor temperature (TEMP), humidity (HUMID), CO$_2$ and sound (SOUND) in every 5 minutes. Real-time data can be uploaded to the Cloud continuously through the Guest WiFi covered on the campus. Figure ~\ref{figb:co2 and temp} shows the indoor temperature and CO$_2$ level in three rooms at 11:00 am on 11 Sep 2019. We can clearly see that the temperature of room 3 is only 12.3 \textcelsius{} and much lower than the comfortable warmth (18 \textcelsius{}) defined by the World Health Organization's standard \cite{world2007housing}, which may negatively affect student learning in class \cite{jiang2019measuring}. Furthermore, CO$2$ levels in room 2 and room 3 are beyond 2000 ppm, which has been proved to have a negative influence on the student cognitive load in the classroom \cite{haverinen2011association,satish2012co2}. Based on previous studies \cite{wargocki2017quantitative}, students may become sleepy and inattentive during the class when the CO$_2$ level is too high.

\subsubsection{Ground Truth: Self-report Survey Instrument Data}

In this study, we choose to use self-report survey to gather subjective measurements of students' in-class engagement. As discussed in Section ~\ref{sec:related work}, the self-report survey is the most common way to measure student engagement as they can reflect students' subjective perceptions explicitly. Instead, measures relying on experience sampling, teacher ratings, interviews or observations have been reported to be easily affected by the external factors. The questionnaire includes 5 items related to behavioural, emotional, and cognitive engagement of the validated \textit{In-class Student Engagement Questionnaires} (ISEQ) \cite{fuller2018development}, which has been proved to be effective for measuring multidimensional engagement compared to the traditional long survey. Similar to \cite{di2018unobtrusive, huynh2018engagemon}, we slightly adapted survey questions from university lectures to high school class context to make the survey easier for students underage to understand. Moreover, for cognitive engagement measurement, we did not use the original question ‘\textit{the activities really helped my learning of this topic}' in \cite{fuller2018development}, considering that some classes in high school do not have in-class activities. Instead, we use the well-accepted item '\textit{I asked myself questions to make sure I understood the class content}' \cite{moore2005conceptualizing}, which is a good reflection of cognitive engagement. 
Table ~\ref{tab:survey} shows the questionnaire used for measuring multidimensional student engagement in class, where item 1,3 and 5 assess the behavioural, emotional and cognitive engagement, item 2 and 4 indicate the behavioural and emotional disaffection \cite{skinner2009motivational,fuller2018development} .

\begin{table*}[]
\caption{Self-report items for measuring in-class engagement in online survey.}
\label{tab:survey}
\begin{tabular}{@{}lll@{}}
\toprule
Questions (please describe your engagement in the last class)                                                                 & Subscales \\ \midrule
1. I paid attention in class.                     \                                                                                                         & Behavioural           \\
2. I pretended to participate in class but actually not.                                                        & Behavioural (-) \\
3. I enjoyed learning new things in class.                                                                      & Emotional   \\
4. I felt discouraged when we worked on something.                                                        & Emotional (-) \\
5. I asked myself questions to make sure I understood the class content.  & Cognitive    \\ \bottomrule
Note: (-) means the reversed score.
\end{tabular}
\end{table*}

\begin{figure}
    \centering
    \subfigure[Question 1]{\includegraphics[width=0.208\textwidth]{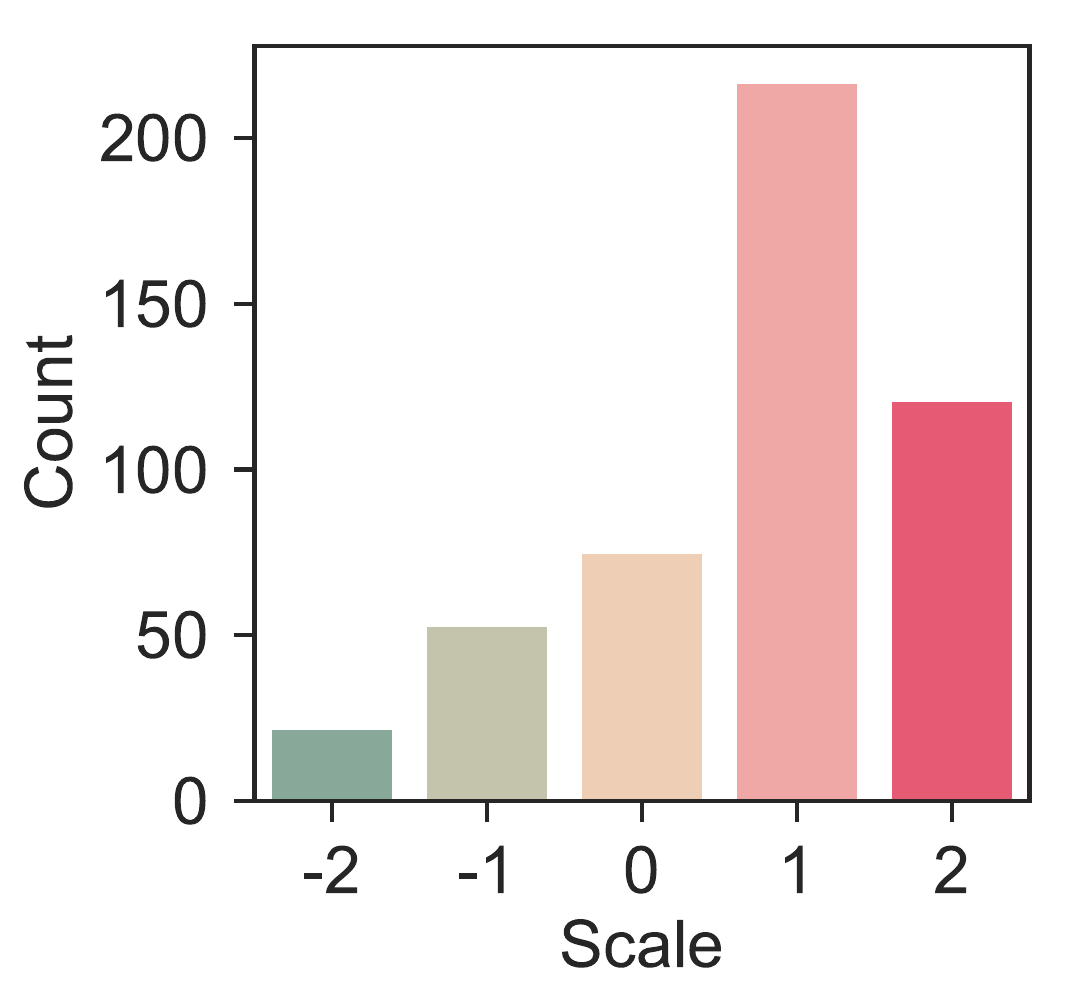}}
    \subfigure[Question 2]{\includegraphics[width=0.193\textwidth]{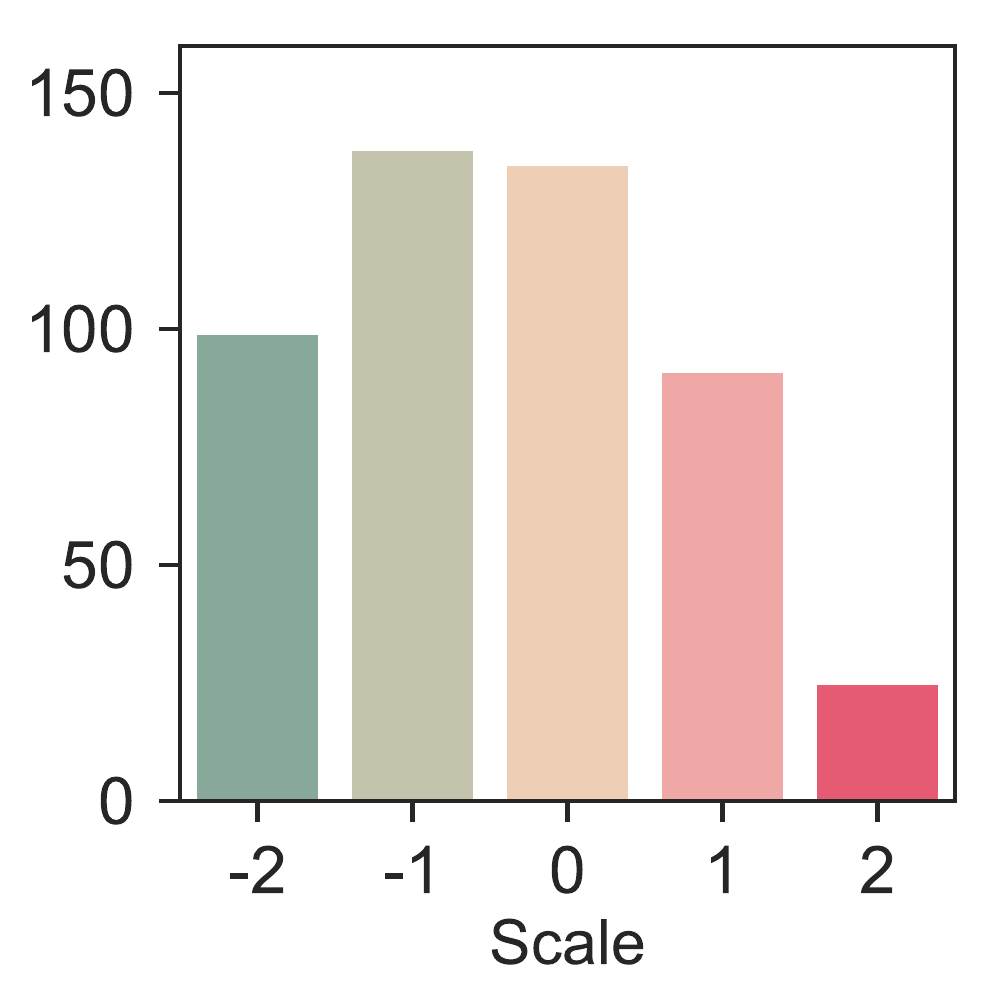}}
    \subfigure[Question 3]{\includegraphics[width=0.193\textwidth]{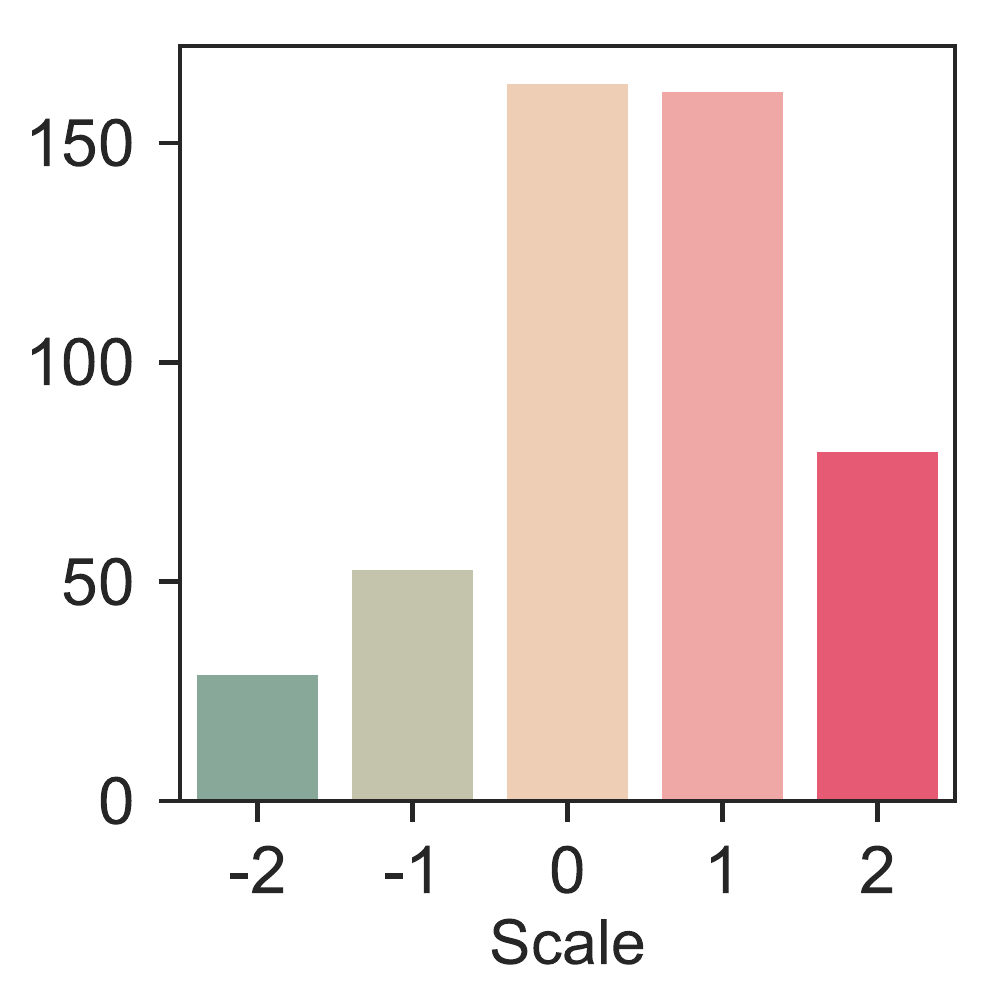}}
    \subfigure[Question 4]{\includegraphics[width=0.193\textwidth]{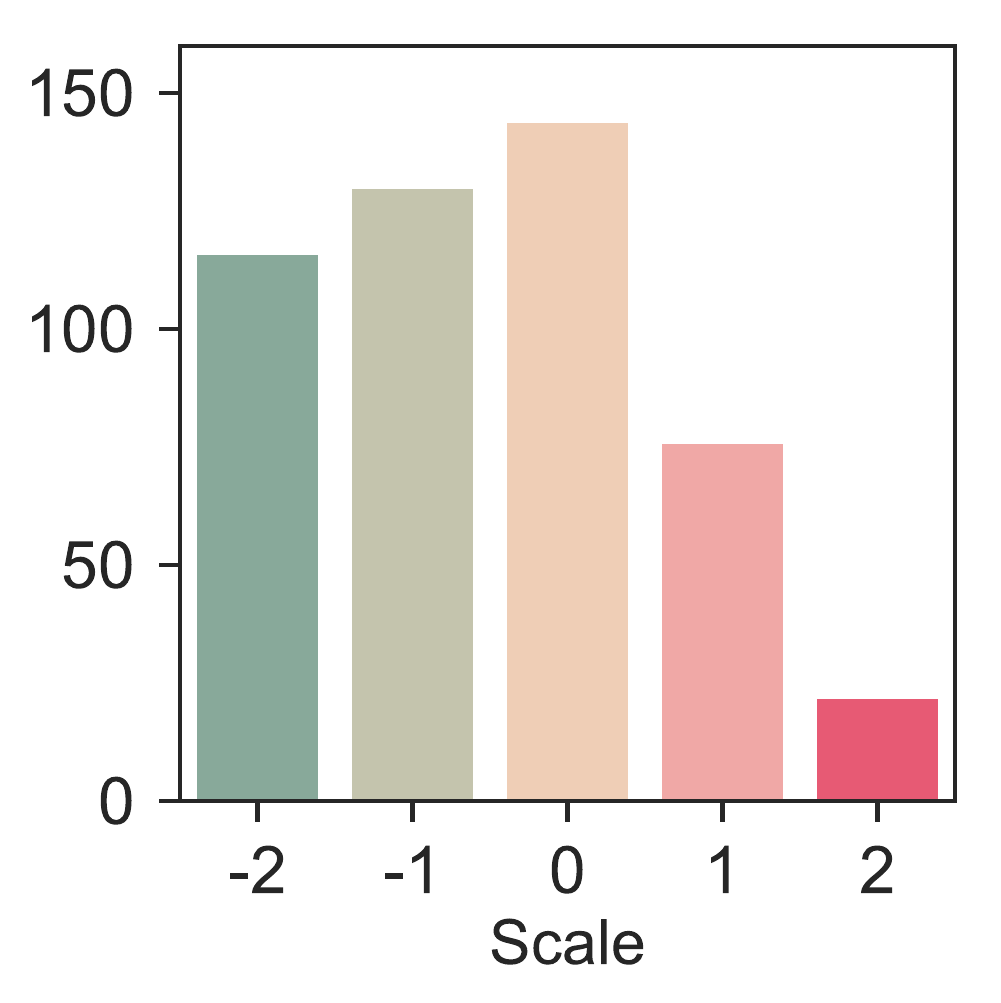}}
    \subfigure[Question 5]{\includegraphics[width=0.193\textwidth]{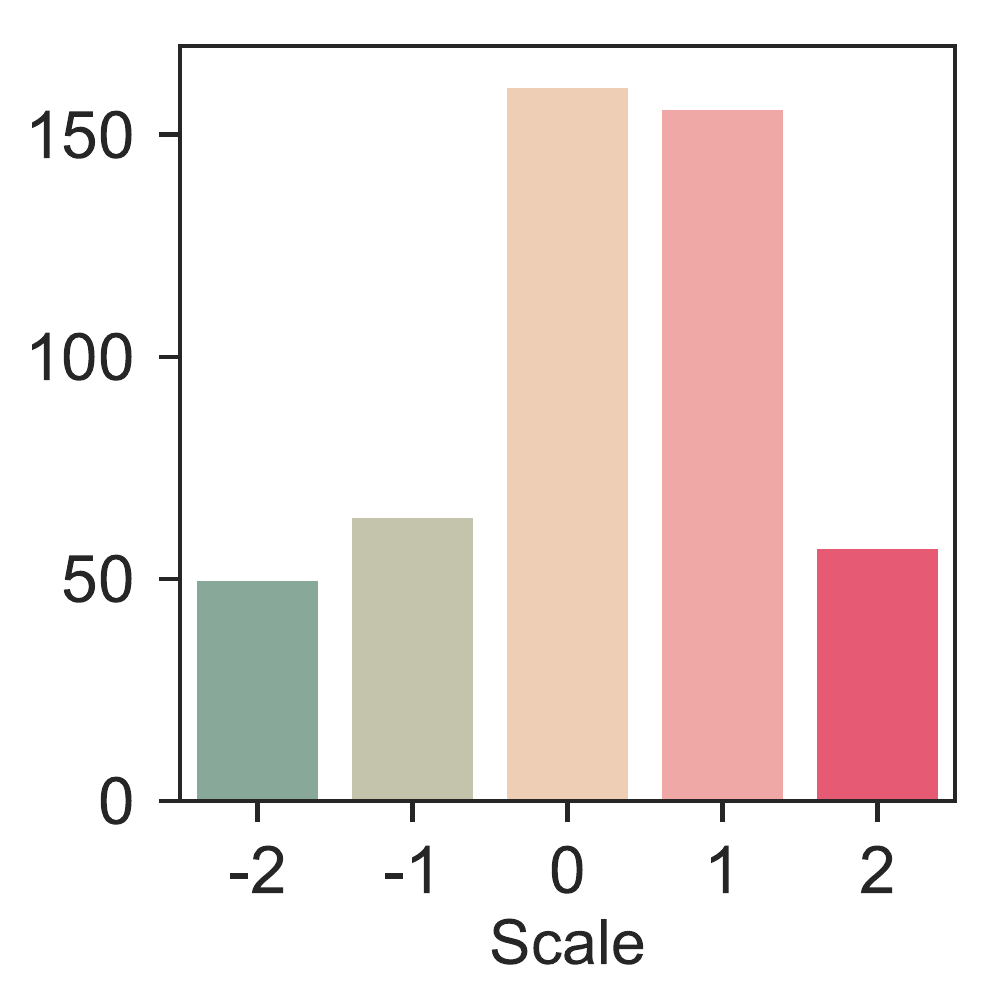}}
    \caption{Histograms of the Answers. The X axis shows the 5-Likert scale from -2 to 2 which means 'strongly disagree' to 'strongly agree'. The Y axis shows the number of the responses that fall into the specific scale.}
    \label{fig:answers}
\end{figure}

In the questionnaire, each item \footnote{In the survey, participants were also asked to report their thermal feelings and mood using the Photographic Affect Meter (PAM) \cite{pollak2011pam}. Nevertheless, this data was not considered in this paper.} is rated with a 5-point Likert-scale from -2 to 2, which indicates 'strongly disagree', 'somewhat disagree', 'neither agree nor disagree', 'somewhat agree' and 'strongly agree'. Figure ~\ref{fig:answers} shows the distribution of responses for each item from total 488 responses. The online self-report survey is constructed with the external tool named \textit{Qualtrics} \footnote{Qualtrics: \url{https://www.qualtrics.com/au/}}. Participants were asked to complete the survey on the public tablets or their digital products with the given survey link generated by \textit{Qualtrics}. 



\subsection{Procedure}
Before the data collection, all wristbands were synchronized with the E4 Manager App from the same laptop to make sure the internal clocks are correct. 1 Netatmo weather station was installed and 1 tablet was put on the teacher desk in each classroom. Students were asked not to unplug the Netatmo stations during the data collection. 

The first two weeks of data collection occurred in early September (winter in the southern hemisphere), and the next two weeks of data collection completed in November (spring in the southern hemisphere). We collected data from two different seasons to build a more robust engagement sensing system. As we know, different seasons usually result in different indoor environments (e.g., indoor temperature, humidity), which may affect students' sweat level (EDA, ST) and activity level (ACC, HRV). If we use the data from one season to build the engagement prediction model, the prediction performance can be greatly reduced in another season due to changes in activity, physiological, and environmental data. Before the data collection, 1 participant was chosen as the representative in each form class, for a total of 3 representatives. 
During the data collection, student participants were distributed with the same wristband (attached with the student ID label) from the representative at 8:50 before the first class started at 9:00. Then at the end of the school day (i.e., 15:35), the representative would remind student participants to hand in wristbands. Student participants were asked to wear the wristband on non-dominating hands and avoid pressing the button or performing any unnecessary movements during class. For teacher participants, they only need to wear the wristband during their classes.

On each school day, student participants were asked to complete the online surveys (either through the public tablets or their own digital devices) at 11:00, 13:25, 15:35 (right after the 2$^{nd}$, 4$^{th}$, 5$^{th}$ class). The length of 2$^{nd}$ and $4^{th}$ class can either be 40 minutes or 80 minutes on the different school day and the 5$^{th}$ class always lasts for 80 minutes. From the class table for Year 10 students, they have the same class schedule on the 1${^{st}}$ week and 3$^{rd}$ week, and another class schedule on the 2$^{nd}$ and 4$^{th}$ week. Each representative would remind student participants to complete online surveys on time.
However, considering that it could be a burden for some participants to complete the survey 3 times a day, we did not urge students to complete the survey for ensuring the quality of survey responses. By the end of the 4$^{th}$ week, we had received 488 valid responses in total and the response rate is 35.3\%. 

As a token of appreciation, the certificate of participation and four movie vouchers were provided to the participants during the 4-week data collection. It is worth noting that participation in this research project is voluntary. Participants are free to withdraw from the project at any stage if they change their minds. Besides, we anonymized all the participants to protect their privacy. 

\section{Data Preprocessing}
\label{sec: preprocessing}

In this section, we first extract class periods based on students' accelerometer data using unsupervised time series segmentation method. Then we introduce the data cleaning process and data pre-processing technique for electrodermal activity, blood volume pulse, accelerometer data, and skin temperature data. 

For data preparation, we only remain the data between 9:00 am to 15:35 pm, which corresponds to the start time of the first class and the end time of the last class. In addition, some students may have several data recording segments during the same day due to the unexpected closure and re-open of the wristband. We drop the data segments less than 15 seconds which is less helpful for extracting useful information. We also discard the data on Tuesday in the last week because students had trip travel and did not have classes on that day.

\subsection{Class Period Segmentation}
As described in Section ~\ref{sec: data collection}, student participants wear wristbands all day along and teachers participants are only asked to wear the wristband at their classes. Participants report their engagement for the 2$^{nd}$, 4$^{th}$, 5$^{th}$ classes of the day during recess time, lunchtime and before going home. Though the scheduled class start/end time is already known, teachers may start/finish the class a bit earlier or later than the scheduled time. The accurate class time is significant for wearable data analysis because participants may have very different physiological/movement patterns between in-class and after-class. For instance, increased activity level after class may lead to a higher value of EDA (due to the higher level of sweat) and variation of accelerometer data. 

\begin{figure}
    \centering
    \includegraphics[width=0.83\textwidth]{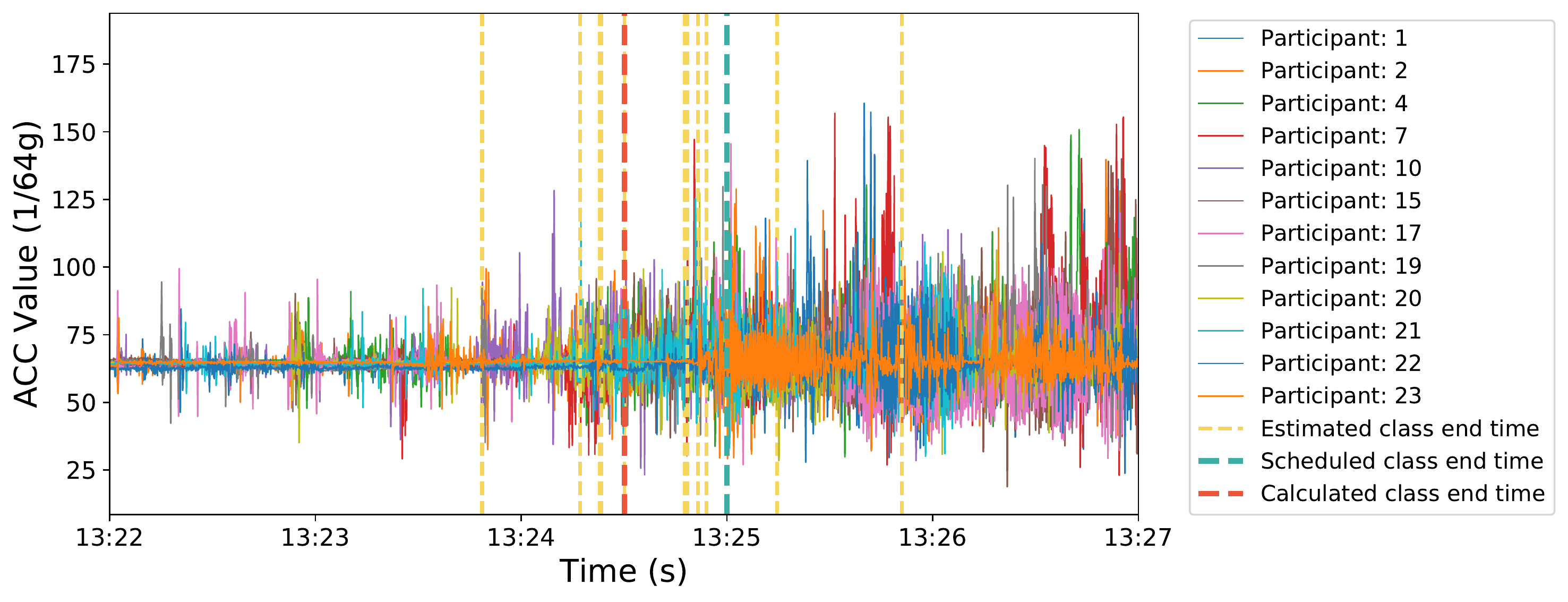}
    \caption{Calculated class end time with ACC data from 12 student participants.}
    \label{fig:igts}
\end{figure}

To get the exact class start/end time for meaningful data analysis, we segment the accelerometer data from student participants based on the assumption that students usually have different activity patterns before/after class. \textit{Information-Gain based Temporal Segmentation} (IGTS) \cite{sadri2017information, deldari2016inferring} is applied on the ACC data to calculate the class start/end time. IGTS is an unsupervised segmentation technique, aims to find the transition times in human activities, which is suitable for dividing the boundary between in-class and out-class \cite{sadri2017information}. Topdown optimization is adopted in the ACC time-series segmentation. To calculate the class boundary, we choose the ACC time-series from 5 minutes before the class to 5 minutes after the class. Take calculating the actual class end time as an example, from Figure ~\ref{fig:igts}, there are 12 participants in a class and the scheduled class end time is 13:25 (green vertical dashed line). Applying IGTS on the ACC data, we can get 12 different estimated class end time from 12 ACC traces. Then, the median time is chosen as the calculated class end time (red vertical dashed line). That is to say, this class finishes early than the scheduled time. We apply IGTS on all the class data and extract the exact class start time and end time for the later data analysis.

\subsection{Data Cleaning}
Before pre-process the collected data, a data cleaning stage needs to be conducted to remove noises from wearable data. As describe in \cite{bakker2011s,di2018unobtrusive,hernandez2014using,gashi2019using,boucsein2012electrodermal}, there are several noises commonly happened in data collection from E4 wristband: (1) flat responses (i.e., 0 micro siemens) due to poor contact between the sensors and the skin. If the contact is not tight enough, the sensor will not measure anything; (2) abrupt signal drops due to the movement of the sensor (e.g., participant bumps the wristband onto the desk); (3) quantization errors. Since EDA sensor records data through the two electrodes, which is more susceptible to noises compared with ACC, PPG and ST sensor, we clean the data set mainly based on the quality of EDA data. 

Firstly, we remove the data when students did not wear wristbands during the whole class or closed off the wristbands unintentionally during the class. Similar to \cite{gashi2019using}, we then discard the signals containing a huge number of flat responses, abrupt signal drops and quantization error as suggested in \cite{hernandez2014using, boucsein2012electrodermal}. Finally, we discard the class data from the student who did not complete the survey. The data cleaning stage leaves us with 331 class data sessions. The final wearable data are gathered from 23 students and 6 teachers in 105 classes. 59 classes are short classes (mean = 39.15 minutes, STD = 1.15 minutes) and 46 classes are long classes with 2 periods (mean = 78.21 minutes, STD = 4.33 minutes).

The data cleaning stage brings to the elimination of 157 class data sessions due to the lack of survey data, which takes up to 32.17\% of the total data with completed surveys. Though the number of eliminated data is considerable, the size of our collected data is comparable and even larger than the previous studies. For instance, Lascio et al. \cite{di2018unobtrusive} used 197 EDA data sessions after a reduction up to 37\%, Gashi et al. \cite{gashi2019using} used 40 presenter-audience EDA pairs with the elimination of 72 pairs. Hernandez et al.  \cite{hernandez2014using} used 51 data sessions with the elimination of 28\% from the original data.

\subsection{Data Pre-processing}
The pre-processing procedure is crucial improve the quality of collected data. For EDA signals, we follow the same pre-processing steps as suggested in \cite{di2018unobtrusive, gashi2019using, hernandez2014using}. (1) Artifacts removal. To mitigate the influence of motion artifacts (MAs), we apply a median filter on  EDA data with a 5-second window as in \cite{di2018unobtrusive}. (2) Decomposition. EDA signal combines a tonic component and a phasic component \cite{boucsein2012electrodermal, cacioppo2007handbook}. The tonic component varies slowly and reflects the general activity of sweat glands influenced by the body and environmental temperatures. The phasic component indicates rapid changes and related to the responses to internal and external stimuli. EDA signals are decomposed with cvxEDA approach \cite{greco2015cvxeda} using convex optimization. (3) Normalization. The amplitude of the EDA signal varies a lot among different people \cite{cacioppo2007handbook} and thus limits the possibility of comparing the signal directly. We normalize the mixed, tonic and phasic EDA values similar to \cite{gashi2019using}.

PPG data, also known as BVP, is provided by the E4 wristband. Similar to \cite{huynh2018engagemon}, we extract IBI signals by detecting the systolic peak of the heartbeat waveform signals from the raw PPG data (window size = 0.75 seconds). Linear interpolation is applied when the heartbeat intervals can not be detected successfully from the low-quality (e.g., motion artifacts) PPG signal. For the ACC data, we calculate the magnitude of 3-axis accelerations as $ \left | a \right | = \sqrt{x^2+y^2+z^2}$. Then a median filter with 0.2 seconds is applied to the magnitude value. Finally, we apply a median filter on the ST data with 0.5 seconds.

\section{Feature Extraction}
\label{sec: data analysis}

We use various sensing devices to infer multidimensional engagement level of high school students. Table ~\ref{tab:feature} summarizes these features. Then, we introduce the computed features and discuss why we explore such sensors and features. 



\begin{table}[]
\caption{Description of the features computed for different sensors}
\label{tab:feature}
\begin{tabular}{@{}lll@{}}
\toprule
\textit{Sensors}      & \textit{Feature name}   & \textit{Description of features}                                       \\ \midrule
\multirow{12}{*}{EDA} & eda/tonic/phasic\_avg   & Average value for the raw, tonic, phasic data                          \\
                      & eda/tonic/phasic\_std   & Standard deviation for the raw, tonic, phasic data                     \\
                      & eda/tonic/phasic\_n\_p  & Number of peaks for the raw, tonic, phasic data                        \\
                      & eda/tonic/phasic\_a\_p  & Mean of peak amplitude for the raw, tonic, phasic data                 \\
                      & eda/tonic/phasic\_auc   & Area under the curve of the raw, tonic, phasic data                    \\
                      & num\_arouse             & Number of arousing moments during the class                            \\
                      & ratio\_arouse           & Ratio of arousing and unarousing moments                               \\
                      & level$_k$              
                      & Ratio of the number of level$_k$ and the length of $S_k$ \\
                      & eda/tonic/phasic\_pcct  & Pearson correlation coefficient with teacher                           \\
                      & eda/tonic/phasic\_pccs* & Pearson correlation coefficient with average value of students         \\
                      & eda/tonic/phasic\_dtwt  & Dynamic time wraping distance with teacher                             \\
                      & eda/tonic/phasic\_dtws* & Dynamic time wraping distance with average value of students           \\ \midrule
\multirow{10}{*}{PPG} & hrv\_bpm                & Average beats per minutes                                              \\
                      & hrv\_meani              & Overall mean of RR intervals (Meani)                                   \\
                      & hrv\_sdnn               & Standard deviation of intervals (SDNN)                                 \\
                      & hrv\_lf\_power          & Absolute power of the low-frequency band (0.04–0.15 Hz)                \\
                      & hrv\_hf\_power          & Absolute power of the high-frequency band (0.15–0.4 Hz)                \\
                      & hrv\_ratio\_lf\_hf      & Ratio of LF-to-HF power                                                \\
                      & hrv\_rmssd              & Root mean square of successive RR interval differences                 \\
                      & hrv\_sdsd               & Standard deviation of successive RR interval differences               \\
                      & hrv\_pnn50              & Percentage of successive interval pairs that differ \textgreater 50 ms \\
                      & hrv\_pnn20              & Percentage of successive interval pairs that differ \textgreater 20 ms \\ \midrule
\multirow{6}{*}{ACC}  & acc\_avg                & Average physical activity intensity during the class                   \\
                      & acc\_std                & Standard deviation of physical activity intensity in class             \\
                      & acc\_dtw\_t             & Dynamic time wraping distance with teacher                             \\
                      & acc\_dtw\_s*            & Dynamic time wraping distance with average value of students           \\
                      & acc\_pcc\_t             & Pearson correlation coefficient with teacher                           \\
                      & acc\_pcc\_s*            & Pearson correlation coefficient with average value of students         \\ \midrule
ST                    & sktemp\_avg/max/min     & Average/maximum/minimum value of skin temperature                      \\ \midrule
CO2                   & mean/max/min\_co2       & Average/maximum/minimum value of CO2                                   \\
TEMP                  & mean/max/min\_temp      & Average/maximum/minimum value of indoor temperature                           \\
HUMID                 & mean/max/min\_co2       & Average/maximum/minimum value of humidity                                 \\
SOUND                 & mean/max/min\_temp      & Average/maximum/minimum value of sound                           \\ \bottomrule
\end{tabular}
\end{table}

\subsection{EDA-based Features}
EDA is a common measure of autonomic nervous system activity, with a long history being used in psychological research \cite{mendes2009assessing}. Recently, EDA measurements have been increasingly explored in affective computing such as the detection of emotion \cite{bakker2011s, canento2011multimodal}, depression \cite{sarchiapone2018association} , and engagement \cite{di2018unobtrusive, hernandez2014using, latulipe2011love}. From EDA data, we extract statistical features such as the standard deviation from EDA (mixed, tonic, phasic) data, which reflects the overall general arousal during the class \cite{di2018unobtrusive}. As suggested in \cite{di2018unobtrusive}, we extract the number of arousing/arousing states, the ratio of arousing states, etc. to show the momentary engagement during the class. The similarity-based method such as Pearson Correlation Coefficient (PCC) \cite{benesty2009pearson} and Dynamic Time Wrapping Distance (DTW) \cite{senin2008dynamic} are used to evaluate the physiological synchrony \cite{palumbo2017interpersonal} of the target student and teacher. Inspired by \cite{ward2018sensing}, we also propose some new features (marked with *) to compute physiological synchrony between the target student and the average values of other students, which has proven to be effective in Table ~\ref{tab:correlation}. 

\subsection{HRV-based Features}
HRV is controlled by the autonomic nervous system (ANS), which can be used to evaluate human emotional arousal and cognitive performance \cite{ahuja2003gsr,appelhans2006heart,choi2017heart,nickel2003sensitivity,luque2013cognitive}. With the help of \textit{HeartPy} \cite{van2019heartpy} toolkit, we compute HRV features from IBI signals extracted from the raw PPG data. As suggested in \cite{mehler_reimer_wang_2011,camm1996heart,shaffer2017overview}, HRV features can be analyzed from time-domain and frequency domain. On the time-domain, we capture features such as the mean/standard deviation of RR intervals (Meani, SDNN) which estimates the overall HRV. We also extract features such as standard deviation/root mean square of successive RR interval differences (SDSD, RMSSD), number/percentage of successive interval pairs that differ larger than 20/50 ms (NN20, NN50, pNN20, pNN50), which describes the momentary change of HRV. On the frequency-domain where parameters are computed by applying Fast Fourier Transform (FFT) to the time series of RR intervals \cite{shaffer2017overview}, we compute the absolute power of the low-frequency band (0.04-0.15 Hz) and high-frequency band (0.15-0.4 Hz). Besides, we compute the ratio of LF-to-HF power which reflects the overall balance of the ANS \cite{nagendra2015cognitive}. 

\subsection{Accelerometer-based Features} 
Student behaviour can be inferred from ACC data, which helps us know more about student participation (e.g., team activities) and engagement level in class \cite{ward2018sensing}. For ACC data, we extract features such as the average physical activity and standard deviation, which describes the  statistical characteristics of the student movement during the class. Inspired by \cite{ward2018sensing}, we propose the movement synchrony features such as the DTW/PCC between the target student and the average values of the other students.

\subsection{Other Features} 

Student learning engagement has been found to be affected by the thermal comfort level of students in the classrooms \cite{jiang2019measuring}, which is influenced by many factors such as indoor temperature, humidity, skin temperature, sound, CO$_2$ level, etc \cite{de2002thermal, world2007housing,gao2020transfer,rahaman2020ambient}. Therefore, statistical features are calculated for indoor temperature, CO$2$, sound and humidity, as the overall estimate of the indoor environment during a class. For ST data, statistical features are extracted to estimate the general arousal of student engagement. According to \cite{kane}, when CO$_2$ level is higher than 1000 ppm, occupants may complain about the drowsiness and poor air, and when CO$_2$ level is higher than 2000 ppm, occupants will feel sleepy, headaches and lose attention. Therefore, the above features are selected to study student engagement.


\section{Prediction Pipeline}
\label{sec: prediction pipeline}

Although engagement prediction is usually regarded as a classification problem, where engagement level can be divided into two or three categories \cite{di2018unobtrusive, huynh2018engagemon} based on specific thresholds, it is not a good practice to determine people's psychological characteristics using classification \cite{gao2019predicting}. In this paper, we choose regression rather than classification for multidimensional engagement prediction. In order to predict multidimensional engagement scores of students, we set up a regression-based pipeline as described below. 

\textbf{Engagement Score}: We assign each student a score for each item in the self-report survey. To achieve this, we first reverse the responses in item 2 and item 4, as shown in Table ~\ref{tab:survey}. Then, we calculated a score based on the average of the 5-point Likert scale for each dimension of engagement and the overall engagement. Then we rescale the calculated score to 1 to 5, representing the engagement level being low to high. Figure ~\ref{fig:boxid} shows the calculated overall engagement score for 23 student participants. To save space, we do not 
display box plots of the distribution of the single-dimensional engagement score. 

\textbf{Regressors}: We adopt LightGBM Regressor \cite{ke2017lightgbm, shao2019flight} to predict self-reported multidimensional engagement scores. As one of the most powerful prediction models, LightGBM is an ensemble method combining a set of weak predictors (i.e., regression trees) to make accurate and reliable predictions. It builds the regression tree vertically (leaf-wise) while other algorithms grow trees horizontally (level-wise).  It will choose the leaf with max delta loss to grow. When growing the same leaves, LightGBM algorithm can reduce more loss than other tree-based algorithms such as GBRT \cite{elith2008working}. 

\textbf{Validation}: It is natural to use cross-validation to train and test prediction models when we are not in a data-rich situation.  The purpose of cross-validation is to estimate the unbiased generalization performance of the prediction model.  However, when using the test set for both model selection (hyperparameter tuning) and model estimation, the test data may be overfitted, and the optimistic bias may occur in the model estimation. Therefore, we adopt the nested cross-validation approach \cite{muller2016introduction} with inner loop cross-validation nested in outer loop cross-validation. The inner loop is used for hyperparameter tuning and feature selection, while the outer loop is responsible for evaluating the performance on the test set. 
In the outer loop, similar to the previous human-centred research \cite{di2018unobtrusive, hernandez2014using}, we first divide the data into $n$ groups, where $n$ represents the number of participants, i.e., $n$=23. Each group contains the data for only one participant. Then we apply \textit{k-fold cross-validation} \cite{wiens2008three}  ($k$=5) and on all student groups. Specifically, data from the same student (group) will not appear in the training and test sets at the same time. In the inner loop,  the remaining data groups are split into $L$ ($L$=3) folds, where each fold serves as a validation set in turn. Then we train (grid search) the hyperparameters on the training set, evaluate them on the validation set, and select the best parameter settings based on the performance recordings over $L$ folds. We use the importance vector generated from LightGBM to reduce the feature dimensionality, which calculates feature importance automatically by averaging the number of times a specific feature used for splitting a branch. Higher values indicate higher feature importance. Top-10 features are selected as the new input features to the LightGBM regressor. The heuristic of choosing 10 features is we find that the prediction error is lowest under this threshold in the experiment.

\textcolor{black}{Similar to \cite{di2018unobtrusive,wang2018sensing}, we also perform leave-one-subject-out (LOSO) \cite{friedman2001elements} validation to evaluate the impact of data from individual participant on the overall prediction error. For both $k$-fold and LOSO validation approaches, we calculate the average performance score (i.e., MAE and RMSE) of the regressor in each iteration.}

\textbf{Baselines and Metrics}: We compare the proposed engagement prediction model with three baselines. The first baseline is the standard linear regressor \cite{seber2012linear}, one of the most widely used regression models. The second baseline takes the average score of each dimension of engagement. The third baseline randomly generates a sample from the distribution of engagement scores and regards it as a predicted value. Similar random baselines have been widely used in previous ubiquitous computing studies such as \cite{wang2018sensing, di2018unobtrusive}. To evaluate the prediction performance of the proposed model, we use the Mean Absolute Error (MAE) and Root Mean Squared Error (RMSE) \cite{chai2014root} metrics.

\section{Results and Discussion}
\label{sec:results}
In this section, we conduct extensive experiments to evaluate the prediction performance of \textit{n-Gage}. We answer the first research question '\textit{Can we measure the multiple dimensions of high school student's learning engagement including emotional, behavioural and cognitive engagement in high schools with sensing data in the wild?}' in Section \ref{sec: results overall}. We answer the second research question '\textit{Can we derive the activity, physiological, and environmental factors contributing to the different dimensions of student learning engagement? If yes, which sensors are the most useful in differentiating each dimension of the  engagement?}'  in Section \ref{sec: results sensors}. We also study how different settings can help improve the performance of \textit{n-Gage}.  Unless otherwise stated, the prediction models are built with LightGBM regressors using all sensors and evaluated by $k$-fold nested cross-validation by default.

\subsection{Overall Prediction Results}
\label{sec: results overall}

\begin{table}[]
\caption{\textcolor{black}{Prediction performance for emotional, cognitive, behavioural, and overall engagement with all sensing data}}
\label{tab:result}
\begin{tabular}{@{}lllllllll@{}}
\toprule
\multirow{2}{*}{\textit{Dimension}} & \multicolumn{4}{c}{\textit{MAE}}         & \multicolumn{4}{|c}{\textit{RMSE}}        \\ \cmidrule(l){2-9} 
                                    & LGBM.       & LR. & Average & Random  &{LGBM.} & LR.      & Average & Random \\ \cmidrule(r){1-9}
\textit{Emotional}                  & \textbf{0.675} & 0.714 & 0.747      & 1.059      & \textbf{0.851} & 0.878 & 0.928      & 1.326       \\
\textit{Cognitive}                  & \textbf{0.906} & 0.921 & 0.977      & 1.288      & \textbf{1.113} & 1.128 & 1.176      & 1.658      \\
\textit{Behavioural}                 & \textbf{0.783} & 0.811 &  0.871      & 1.235      & \textbf{0.960} & 0.980& 1.135       & 1.540      \\\hline
\textit{Overall}                    & \textbf{0.602} & 0.614 & 0.641      & 0.891      & \textbf{0.753} & 0.769 & 0.792      & 1.125      \\ \bottomrule
\end{tabular}
\end{table}

We first evaluate the overall prediction results for \textit{n-Gage} with all sensors available. Table ~\ref{tab:result} displays MAE and RMSE scores of \textit{n-Gage}'s engagement regression in different dimensions. In particular, the overall engagement is calculated by the average of engagement scores from all questions related to the engagement, which is commonly used in previous engagement studies \cite{huynh2018engagemon,di2018unobtrusive, fredricks2004school}. From Table ~\ref{tab:result}, we can see that in terms of MAE and RMSE, \textit{n-Gage} achieves higher prediction performance for all dimensions of engagement than all baselines, demonstrating its potential for multidimensional engagement prediction. 

Notably, among each dimension of engagement, \textit{n-Gage} works best on predicting emotional engagement. The emotional engagement regression model obtain 0.675 of MAE and 0.851 of RMSE, which is lower than 0.384 (36.26\%) and 0.475 (35.82\%) of the random baseline. The reasons why \textit{n-Gage} predicts emotional engagement best are possibly two-fold: (1) compared with cognitive and behavioural engagement, emotional engagement is most suitable for evaluation through self-report surveys \cite{fredricks2004school}, resulting to a more realistic and stable student emotional engagement measurement (ground truth). (2) emotional engagement is more easily detected by sensors (e.g., EDA and PPG) as it reflects the degree of emotional arousal, thereby producing fluctuations in physiological signals \cite{di2018unobtrusive, king2019micro, bakker2011s}.

\begin{figure}[]
    \centering
    \includegraphics[width=0.8\textwidth]{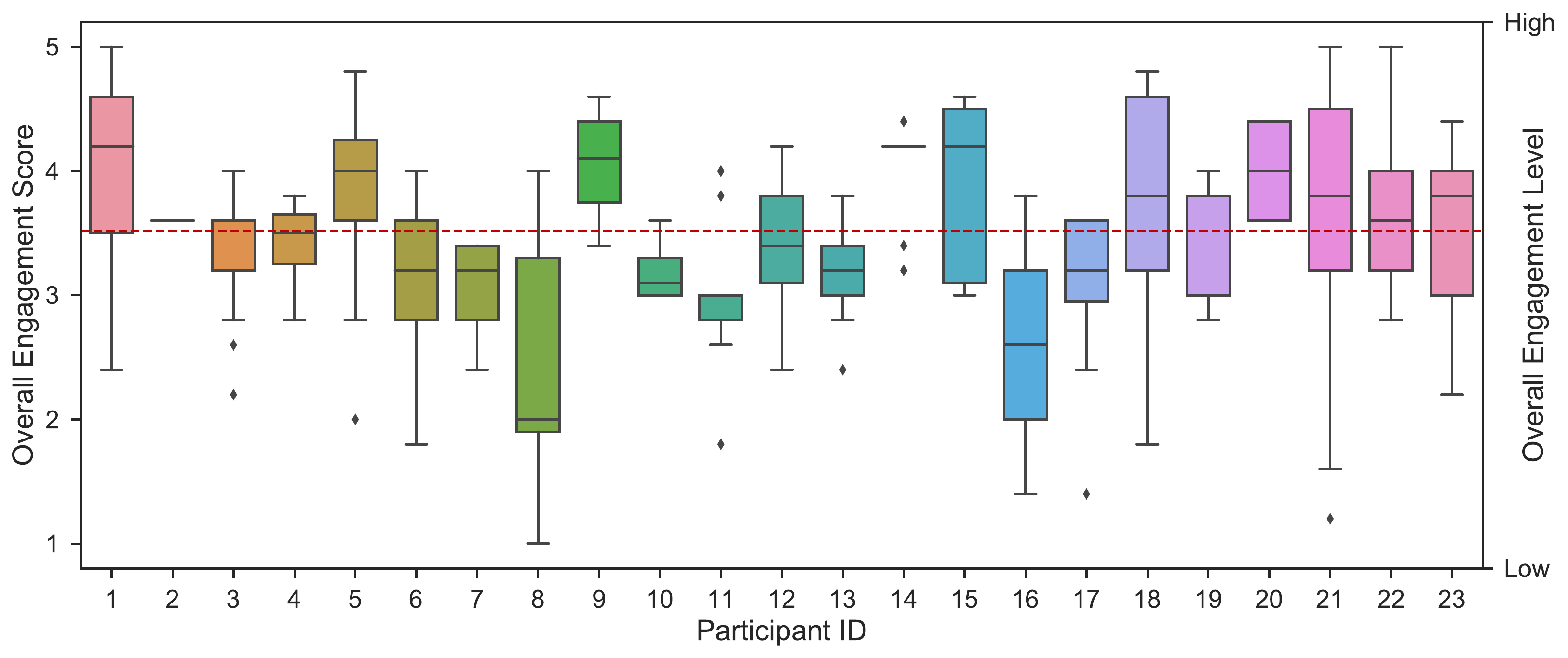}
    \caption{\textcolor{black}{Box plot of the overall engagement scores for 23 student participants. The red dashed line represents the average score for all participants. The participant ID shown in the figure is randomly generated to maintain the privacy of participants.}}
    \label{fig:boxid}
\end{figure}
\begin{figure}[]
    \centering
    \includegraphics[width=0.79\textwidth]{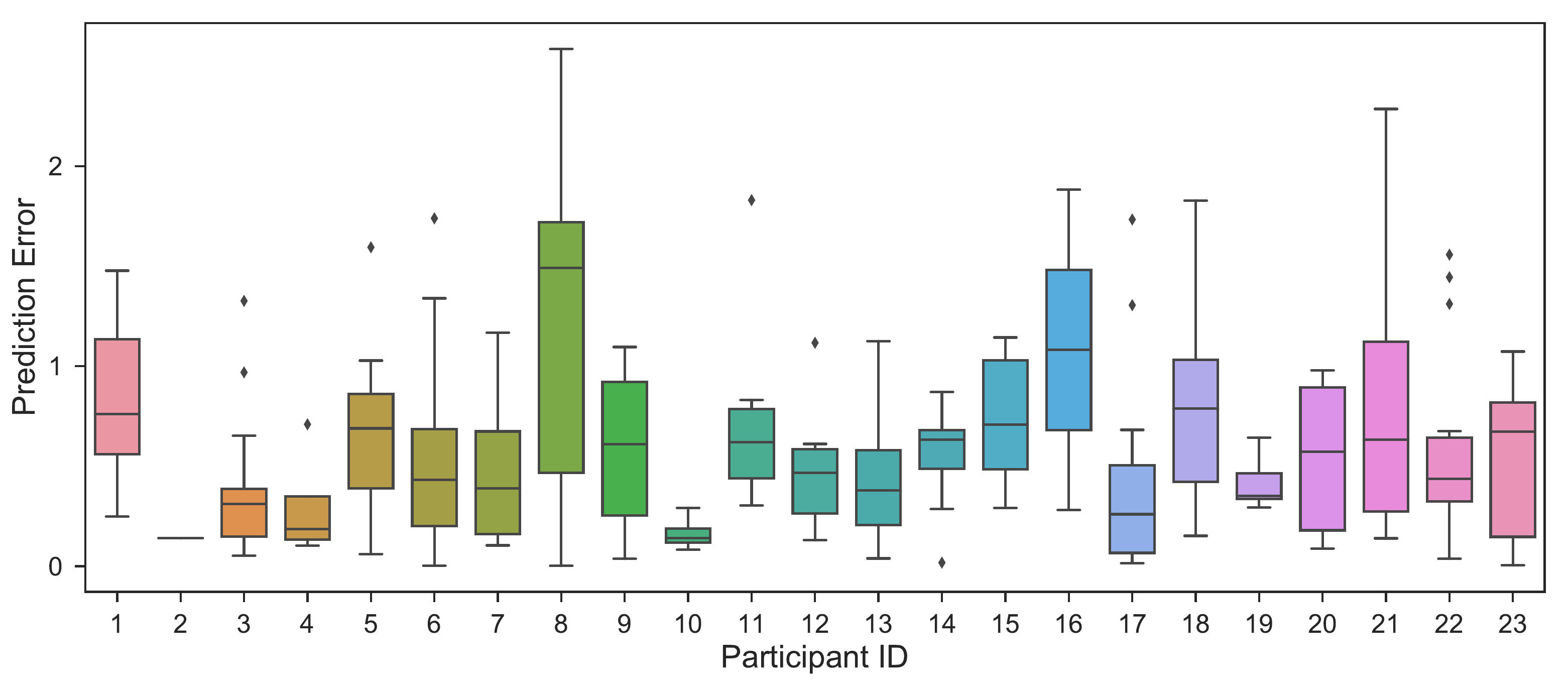}
    \caption{\textcolor{black}{Prediction error for overall engagement scores for 23 student participants}}
    \label{fig:boxerror}
\end{figure}

Although the MAE of cognitive engagement regression is higher than other models, it is still lower than random baseline of 0.382 (29.66\%) in MAE and 0.545 (32.87\%) in RMSE. The possible reason is that cognitive engagement is more challenging to be assessed by the wearable and indoor sensors than electroencephalography (EEG) sensors \cite{berka2007eeg}. By contrast, \textit{n-Gage} has the lowest prediction error of 0.602 in MAE and 0.753 in RMSE in overall engagement assessment. According to the education research \cite{fredricks2004school,fredricks2012measurement}, although the multidimensional concept of engagement has been well accepted, the definitions of three dimensions of engagement vary with considerable overlap across components. Therefore, the overall engagement is easier to be evaluated and predicted than the single-dimensional engagement.

\textcolor{black}{We also compare the prediction results with when a standard linear regressor is learned. From Table ~\ref{tab:result}, 
the linear regression model has much higher prediction performance than both average and random baseline models (e.g., 31.09\% lower than random baseline model in MAE for overall engagement prediction), indicating the effectiveness of extracted features in engagement prediction. However, the performance of linear regressors is not comparable to the LightGBM in all dimensions. This is because LightGBM has a good ability to capture non-linear feature-target relationships which is more flexible than simple linear regressors. To summarize, we believe that the performance of \textit{n-Gage} is benefited from both the extracted features and powerful non-linear mapping provided by the LightGBM.}


\textcolor{black}{We then discuss the impact of data from the individual participant on the overall prediction error. We train and test the regressors using the LOSO validation approach which enables us to evaluate the ability of models to accurately predict a new participant not included in the training set. Figure ~\ref{fig:boxerror} shows the boxplot of absolute prediction error per participant. Interestingly, each participant has a very different error distribution. For instance, participants 8 and 16 have the highest median value (1.492 and 1.082) and standard deviation (0.801 and 1.132) of prediction errors. From Figure ~\ref{fig:boxid}, we observe that both participants have a much lower engagement level than the others. Since the regression model is built on the data from all the other participants, it does not work well when the participant (testing set) has a different distribution from the training set. The potential solution is to build participant-wise or groupwise prediction models, as introduced in \cite{pal2019applications}. In conclusion, we believe the prediction errors come from both the specific participants and overall prediction bias. We will further investigate this issue in future research.}





\subsection{Impact of Sensor Combinations}
\label{sec: results sensors}

\begin{table}[]
\caption{The most influential features on multidimensional engagement.}
\label{tab:correlation}
\begin{tabular}{ccc}
\toprule
\textit{Engagement}                  & \multicolumn{1}{l}{\textit{Association}} & \textit{Most influential features}                  \\ \midrule
\multirow{2}{*}{\textit{Emotional Engagement}}  & (+)                                      & acc\_pcc\_s, tonic\_a\_p, eda\_pcc\_s    \\
                                     & (-)                                      & \textbf{acc\_avg*}, \textbf{sktemp\_avg*}, eda\_dtw\_t       \\\hline
\multirow{2}{*}{\textit{Cognitive Engagement}}  & (+)                                      & \textbf{intemp\_min*}, level\_1, hrv\_ratio\_lf\_hf 

\\
                                     & (-)                                      & \textbf{acc\_pcc\_s*}, co2\_max, acc\_std           \\\hline
\multirow{2}{*}{\textit{Behavioural Engagement}} & (+)                                      & acc\_std, acc\_pcc\_s, eda\_pcc\_avg        \\
                                     & (-)                                      & \textbf{sktemp\_avg*}, \textbf{acc\_pcc\_t*}, acc\_dtw\_t    \\\hline
\multirow{2}{*}{\textit{Overall Engagement}}    & (+)                                      & level\_1, tonic\_a\_p, intemp\_max              \\
                                     & (-)                                      & \textbf{acc\_dtw\_t*}, sktemp\_avg, acc\_avg       \\ \bottomrule 
                              \addlinespace[0.1cm]       
                              * indicates p-value < 0.01. 
\end{tabular}
\end{table}

We will explore the physiological, activity and environmental factors contributing to the different dimensions of student engagement. We compute the Pearson Correlation Coefficient (PCC) between the extracted features and multiple dimensions of engagement, and then list the three most influential features in Table ~\ref{tab:correlation}. We find many EDA features related to the peaks of tonic EDA signals and physiological synchrony are related to the multidimensional engagement. In previous research, EDA features are generally considered as a good indicator of physiological arousal (e.g., emotional and cognitive states) \cite{critchley2002electrodermal, boucsein2012electrodermal}, which have been explored in the detection of engagement \cite{di2018unobtrusive, hernandez2014using, latulipe2011love}. For the HRV features (e.g., 'hrv\_ratio\_lf\_hf'), they are shown to be correlated with cognitive engagement as HRV is an autonomically dependent variable and has been used to predict student engagement in \cite{monkaresi2016automated}. Similar to EDA and HRV features, we notice that the average skin temperature ('sktemp\_avg') are negatively correlated with engagement, as ST reflects the sympathetic nervous activity and attention states \cite{andreassi2010psychophysiology} which has been used for mind-wandering prediction \cite{blanchard2014automated} and stress detection \cite{herborn2015skin}. 

For activity factors, it is interesting to find that many ACC features are highly correlated with engagement. Accelerometer is a popular and powerful sensor for quantifying human behavioural patterns \cite{wang2014studentlife, gao2019predicting}. ACC features have been utilised to sense audience engagement using interpersonal movement synchrony \cite{ward2018sensing}. In the experiment, we observe that the average physical intensity during class is highly negatively correlated with emotional engagement. This leads us to believe that when students are negatively engaged, they tend to perform more physical movements in the class. As for environmental factors, we find that the maximal CO$_2$ level is negatively associated with cognitive engagement, while the indoor temperature is positively associated with engagement. This may be because CO$_2$ has a negative impact on people's cognitive load \cite{haverinen2011association,satish2012co2}, and then affects student cognitive engagement. This result highlights the need to ventilate the classroom timely to keep students engaged. Interestingly, we notice that the maximal indoor temperature in the class is positively correlated with overall engagement. One possible explanation is that during the data collection period (winter and spring), the indoor temperature is low and moderately higher indoor temperature makes students feel thermally comfortable \cite{world2007housing} and therefore more engaged in learning \cite{jiang2019measuring}.



\begin{table}[]
\caption{Summary of the Prediction performance of multidimensional engagement using different sensor combinations. $\mathcal{X}_1$ indicates all the wearable data including EDA, HRV, ACC and ST data, and $\mathcal{X}_2$ means the indoor environmental data including CO$_2$ and temperature data.}
\label{tab:feature sets}
\begin{tabular}{@{}lllll@{}}
\toprule
\multicolumn{1}{c}{\multirow{2}{*}{\textit{Data source}}}                      & \multicolumn{4}{c}{\textit{MAE/RMSE}}                                                  \\ \cmidrule(l){2-5} 
\multicolumn{1}{c}{}                                                           & \textit{Emotional} & \textit{Cognitive}  & \textit{Behavioural}  & \textit{Overall}     \\ \cmidrule(r){1-5}
\textit{EDA}                                                                   & 0.697/0.877        & 0.948/1.149         & 0.851/1.019          & 0.637/0.800          \\
\textit{HRV}                                                                   & 0.714/0.901       & 0.940/1.140         & 0.833/1.002         & 0.659/0.812          \\
\hline
\textit{EDA+HRV}                                                              & 0.699/0.875        & 0.949/1.151         & 0.841/0.989          & 0.621/0.783          \\
\textit{EDA+ACC}                                                              & 0.679/0.860        & 0.914/1.124         & 0.816/0.987          & 0.626/0.789          \\
\textit{HRV+ACC}                                                              & 0.691/0.875        & 0.910/1.125         & 0.809/0.979          & 0.641/0.796          \\\hline
\textit{EDA+HRV+ACC}                                                              & 0.679/0.860        & 0.909/1.122         & 0.800/0.965          & 0.620/0.778          \\
\textit{$\mathcal{X}_1$*}                                                     & \textbf{0.673}/0.851        & 0.910/1.126         & 0.811/0.980          & 0.619/0.775          \\\hline
\textit{\begin{tabular}[c]{@{}l@{}} $\mathcal{X}_1$+ $\mathcal{X}_2$* (all)
\end{tabular}} & 0.675/0.851       & \textbf{0.906/1.113} & \textbf{0.783/0.960} & \textbf{0.602/0.753} \\ \bottomrule
\addlinespace[0.1cm]
\multicolumn{5}{l}{* indicates the proposed combination of features for engagement prediction.}
\end{tabular}
\end{table}

Then we investigate the most useful sensors in predicting each dimension of student engagement and explore the performance of \textit{n-Gage} when only a set of sensors available. In this research, we use E4 wristbands and Netatmo indoor weather stations for student engagement assessment. However, when other schools want to generalize the system for automatic engagement measurement, it is likely that only a few sensors available considering the types of wearables and installation of indoor weather stations. In this experiment, we use different combinations of sensors as shown in Table ~\ref{tab:feature sets} to train the regressors, where $\mathcal{X}_1$ indicates all the wearable sensors including EDA, HRV, ACC and ST, and $\mathcal{X}_2$ represents all the environmental sensors containing CO$_2$, TEMP, HUMID and SOUND sensors. Besides, we predict student engagement using only EDA as in \cite{di2018unobtrusive}, single PPG (HRV) as in \cite{fuller2018development}, and EDA+HRV as in \cite{huynh2018engagemon}. Since accelerometers are naturally available in wearables and have been used for engagement measurement \cite{ward2018sensing}, we add ACC to the above sensor combinations for the first time. Then we utilise all wearable sensors and indoor sensors for more accurate engagement prediction.

For each sensor combination, we use nested cross-validation to train and test the regressors as described in Section ~\ref{sec: prediction pipeline}, to achieve optimal feature selection and parameter tuning. Table ~\ref{tab:feature sets} displays the regression result with different sensor combinations. Different combinations are useful for different dimensions of engagement. For instance, a single EDA sensor works well for emotional engagement prediction while less useful in predicting behavioural engagement unless involving ACC together. This is reasonable because EDA is a reflection of emotional arousal, while ACC is capable of quantifying human behavioural patterns \cite{wang2014studentlife, gao2019predicting}. On the other hand, the combination of EDA and HRV sensors has similar prediction performance compared to using a single EDA sensor, which is consistent with the fact that not many HRV features are highly correlated with engagement. When there is no EDA sensor (especially in commercial off-the-shelf smart wristbands), the HRV+ACC combination can achieve similar prediction performance on cognitive and behavioural engagement compared to EDA+HRV+ACC. 

Meanwhile, it can be observed that the combination of all wearable sensors ($\mathcal{X}_1$) has the lowest prediction error for emotional engagement. When considering wearable sensors ($\mathcal{X}_1$) with indoor sensors ($\mathcal{X}_2$), \textit{n-Gage} can achieve the best performance on the behavioural, cognitive and overall engagement, and has similar prediction performance in emotional engagement with $\mathcal{X}_1$. The underlying reason is that CO$_2$ and indoor temperature mainly affect students' cognition load and behavioural patterns. For example, students may lose attention (related to behavioural engagement), sleepy (related to cognitive engagement) \cite{kane} during class when the CO$_2$ level is too high (e.g., larger than 2000 ppm), but this does not necessarily mean that students do not like the class (related to emotional engagement). The above results illustrate the importance of taking indoor environmental changes into account for student engagement prediction and creating the optimal environment to keep students engaged in class.  

\subsection{Impact of Class Subjects}

\begin{table}[]
\caption{Multidimensional Engagement Regression Result for Different Subjects}
\label{tab: result for different subjects}
\begin{tabular}{@{}lllll@{}}
\toprule
\multirow{2}{*}{Subject} & \multicolumn{4}{c}{MAE/RMSE}                          \\ \cmidrule(l){2-5} 
                         & Emotional   & Cognitive   & Behavioural  & Overall     \\ \midrule
Maths                    & 0.686/0.841 & 0.841/0.965 & 0.750/0.891 & 0.603/0.738 \\
English                  & 0.609/0.779 & 0.893/1.010 & 0.694/0.819 & 0.510/0.629 \\
Language                 & 0.645/0.814 & 0.829/0.903 & 0.799/0.900 & 0.593/0.758 \\
Science                  & 0.646/0.829 & 0.895/0.941 & 0.758/0.856 & 0.575/0.720 \\
Politics                 & 0.674/0.835 & 0.947/1.057 & 0.660/0.731 & 0.525/0.671 \\ \hline
Average                  & 0.652/0.820 & 0.881/0.975 & 0.732/0.839 & 0.561/0.703 \\ \bottomrule
\end{tabular}
\end{table}

Now, we investigate whether considering different school subjects could improve the prediction performance of \textit{n-Gage}. Our assumption here is that different subjects may lead to different learning requirements, thinking styles and emotional preferences. Then, student engagement levels and physiological status may be affected accordingly. 

\begin{figure}[]
	\centering
	\subfigure[MAE Score]{\includegraphics[width=0.38\textwidth]{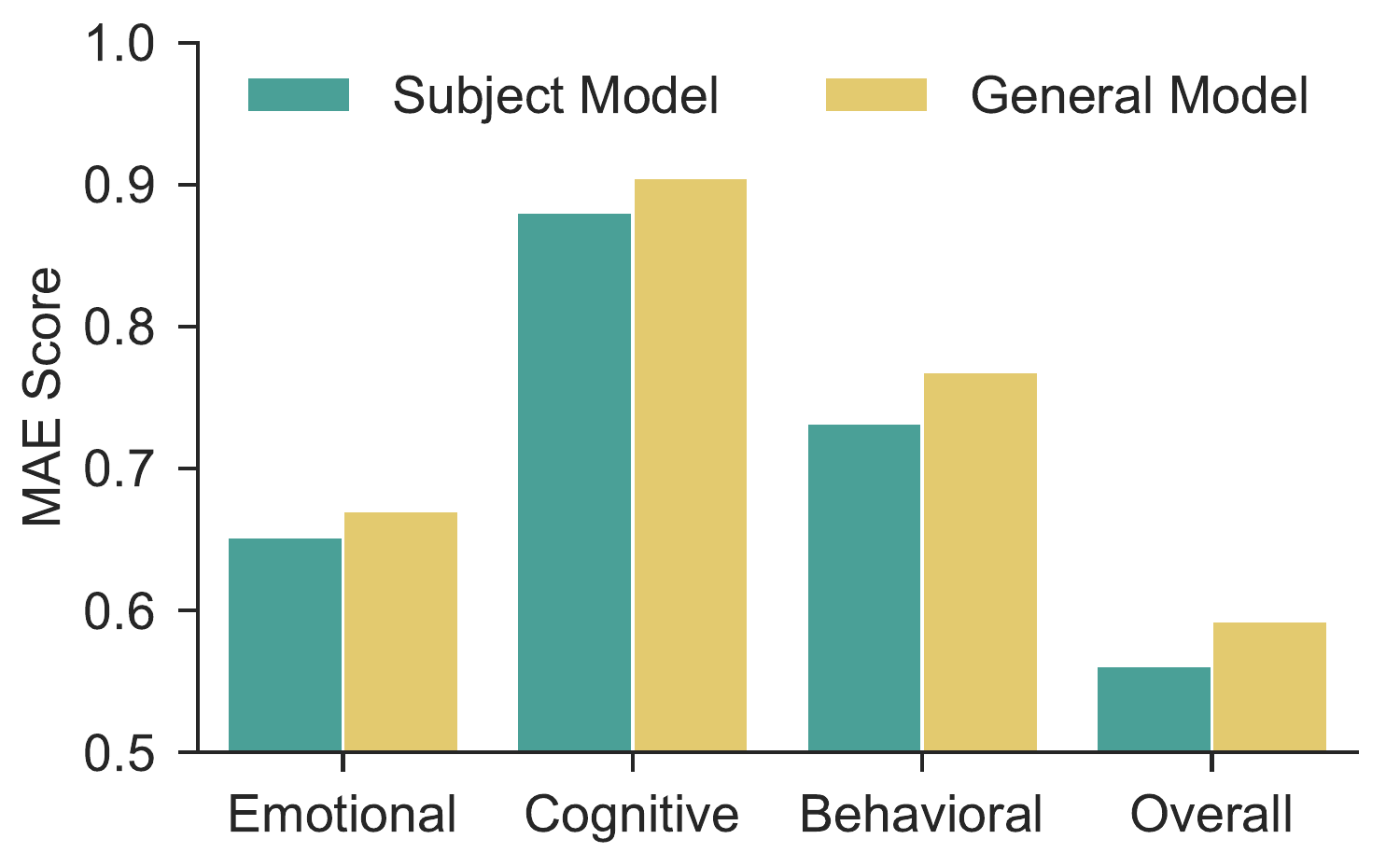}}
	\hspace{0.4cm}
	\subfigure[RMSE Score]{\includegraphics[width=0.38\textwidth]{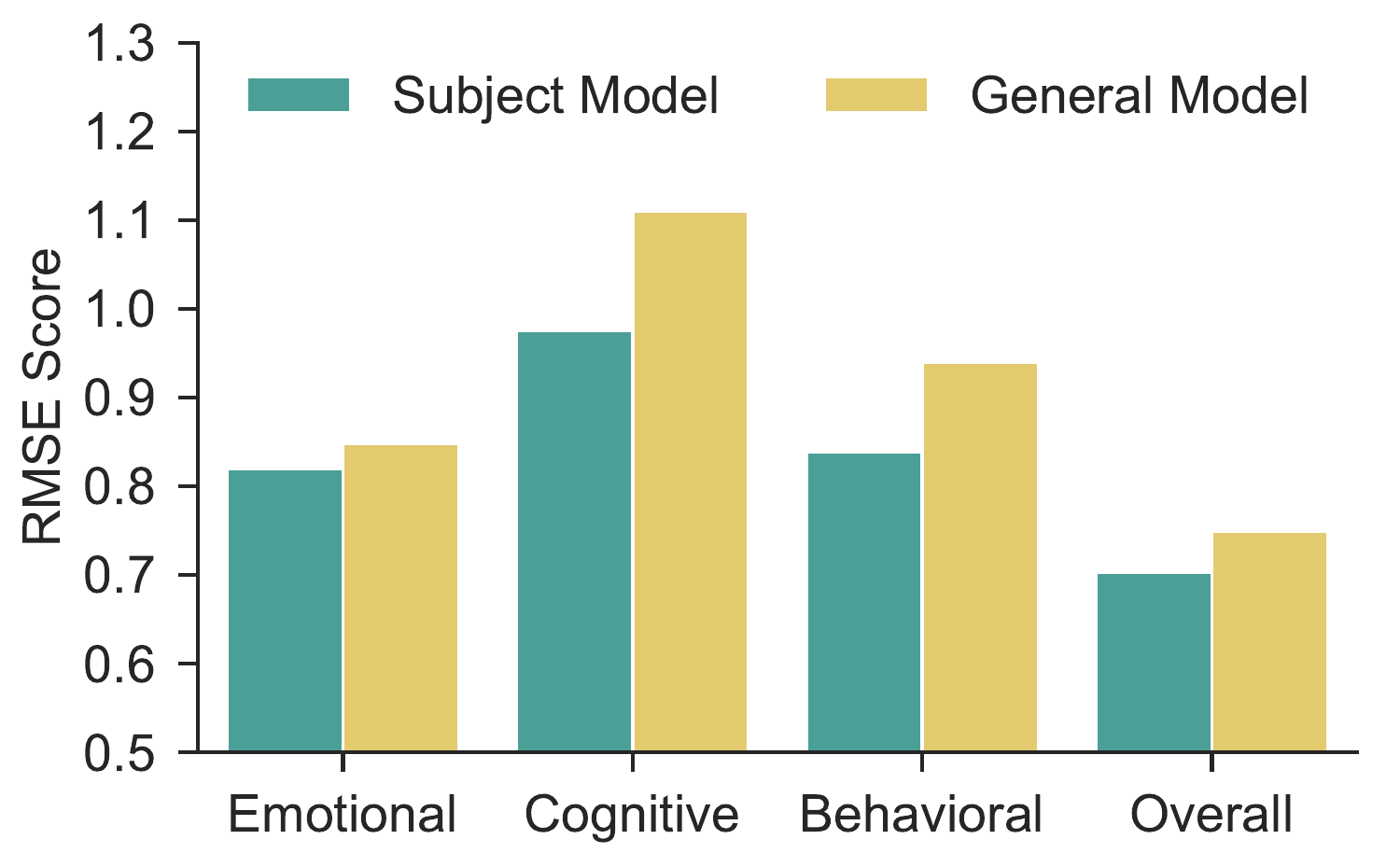}}
    \caption{Prediction performance for the average subject model and general model}
    \label{fig:subject mae mse}
\end{figure}

To validate this hypothesis, we establish regression models for each subject (i.e., Language, Maths, Science, English, PE, Politics, Health, Chapel) to isolate differences in class subjects and engagement assessment. Table ~\ref{tab: result for different subjects} summarizes MAE and RMSE scores of the regressors over different subjects. We do not consider the Health, Chapel and PE classes because the number of survey responses are limited (less than 30) in those classes which may affect the prediction performance. We also compare the average prediction performance of 5 regression models (i.e., Maths, English, Language, Science, Politics) with the general regressor model in Figure ~\ref{fig:subject mae mse}. The results indicate that, compared with building the general regression model including all subjects, building regression models by school subjects can significantly improve the prediction performance.

\begin{figure}
    \centering
    \includegraphics[width=0.81\textwidth]{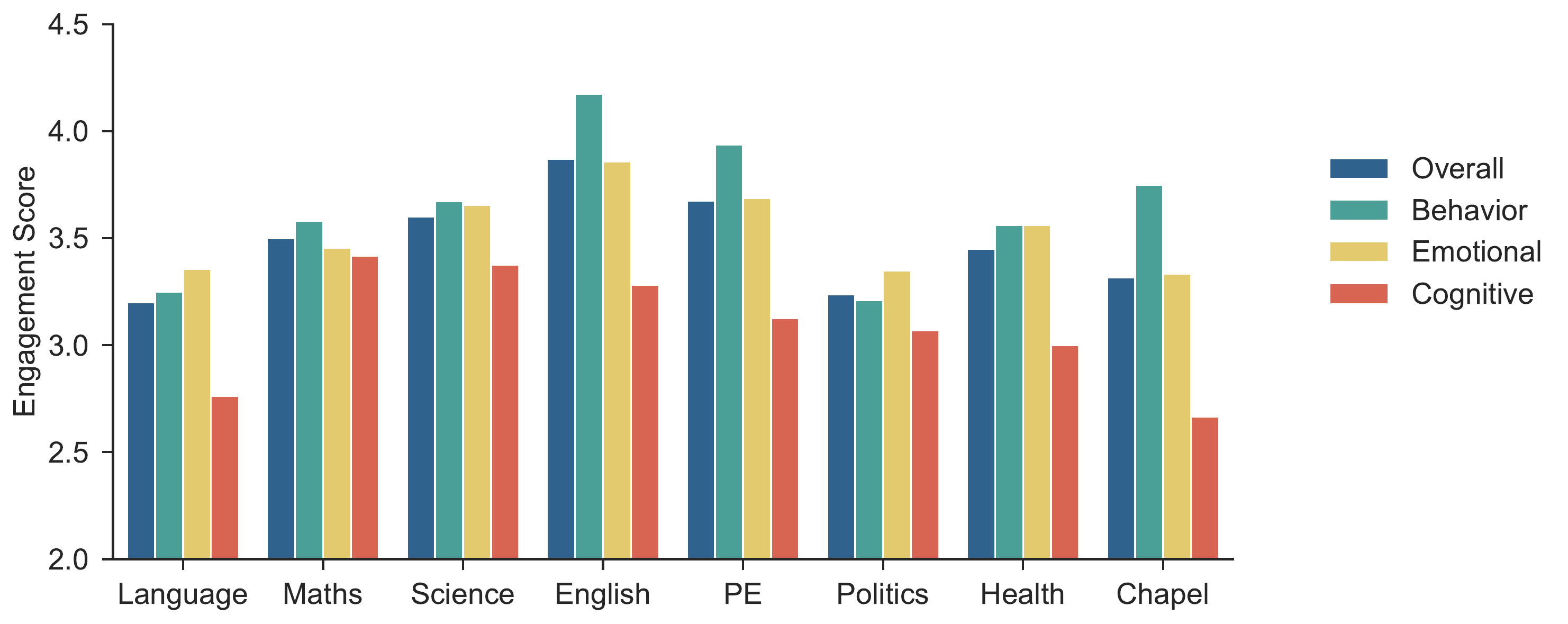}
    \caption{Engagement scores on different subjects}
    \label{fig:dis_subject}
\end{figure}

To better understand the underlying cause behind the improved regression performance, we review the self-reported engagement scores. Figure ~\ref{fig:dis_subject} shows that students have very different multidimensional engagement scores among different subjects. For instance, while students have the highest behavioural and emotional engagement score in English class, they have the highest cognitive engagement score in Maths class. The possible reason is that students enjoy English classes most and thus like to follow the rules from English teachers. Due to the fact that Math know-how is cumulative and usually contains complex concepts, students may tend to put more effort to comprehend the contents in Maths class, thus leading to a high cognition engagement score. Overall, these observations serve as evidence that building models for each subject can lead to significantly improved prediction performance.


\subsection{Discussion}

\begin{figure}
	\centering
	\subfigure[Class time]{\includegraphics[width=0.4\textwidth]{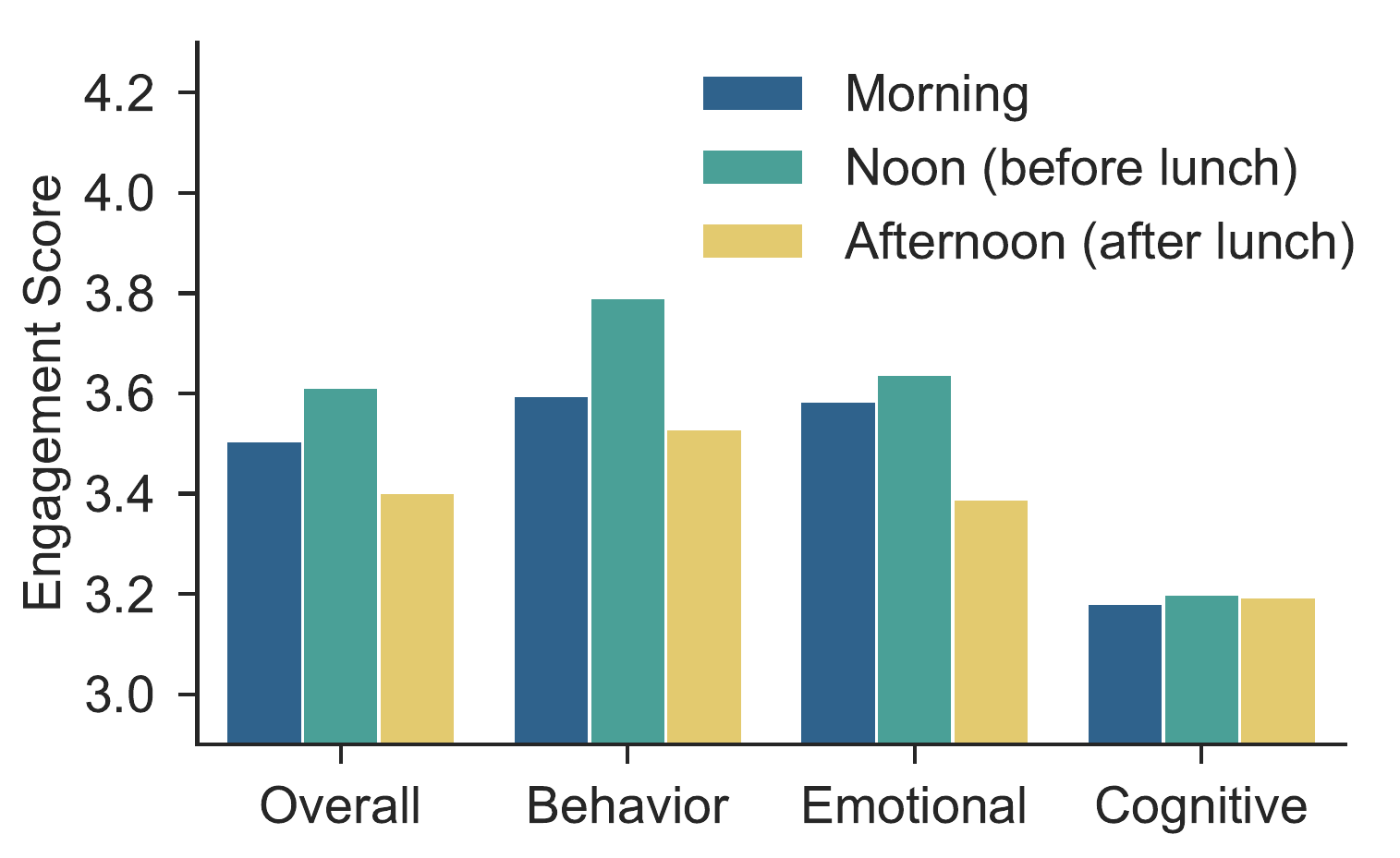}
	\label{fig: class time}
	}
	\hspace{0.4cm}
	\subfigure[Thermal preference]{\includegraphics[width=0.4\textwidth]{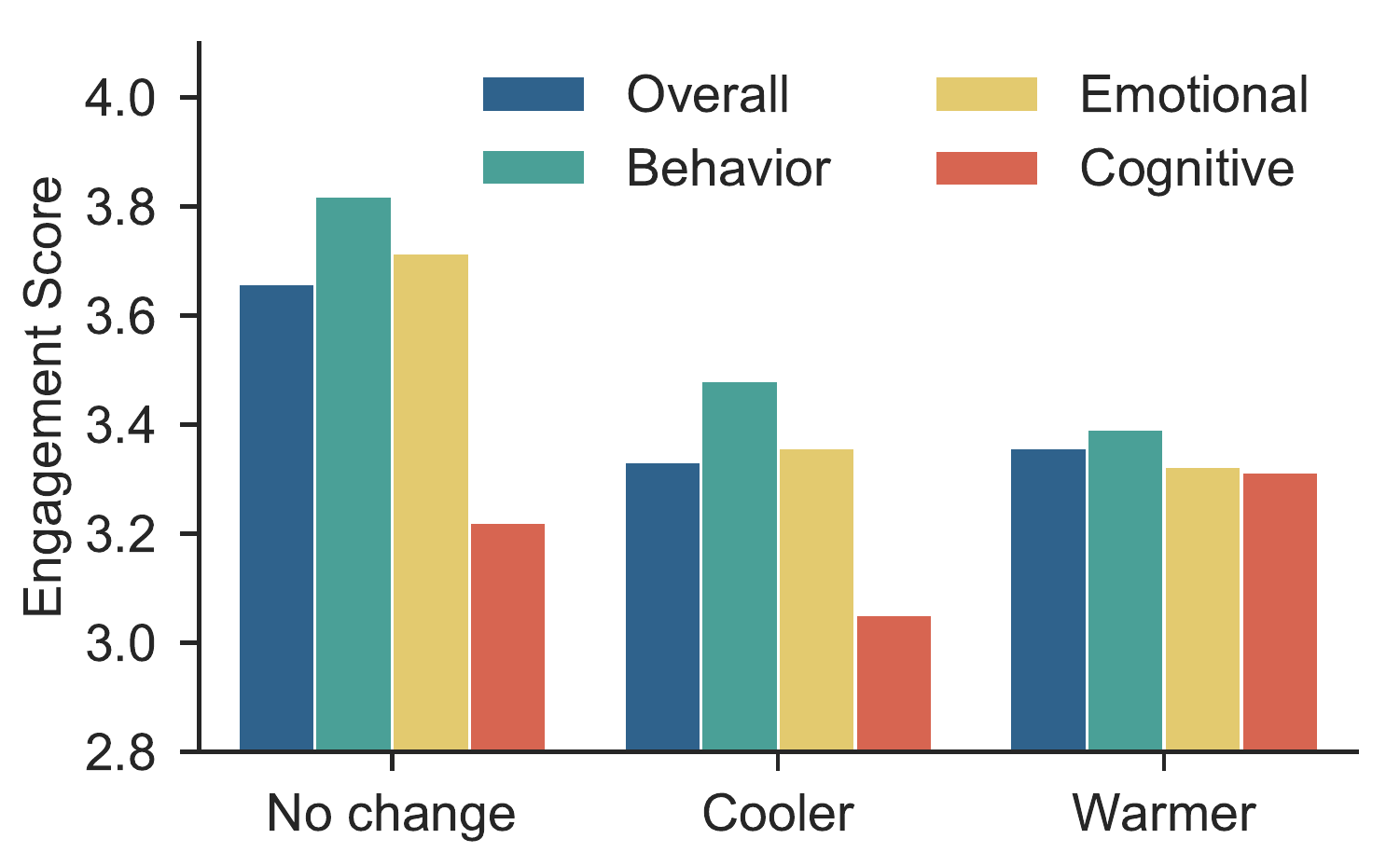}\label{fig: thermal comfort}}
    \caption{Engagement scores with different class time and thermal comfort}
    \label{fig: class time and thermal comfort}
\end{figure}

We have shown that it is possible to infer multidimensional student engagement by using multiple wearable and environmental sensors. Meantime, we will present the following interesting discussion points.
\begin{itemize}
    \item \textbf{Engagement and class time}. A preliminary study is conducted to investigate the correlation between self-reported student engagement and class time during the school day. Figure ~\ref{fig: class time} shows the average engagement scores for the different class time (morning, noon and afternoon). Overall, we observe that classes in the noon show higher engagement levels in all dimensions. Classes in the afternoon (after lunch) have the lowest engagement score, especially in the behavioural and emotional dimensions. Particularly, it is interesting to notice that students have a much higher behavioural and emotional engagement level than the cognition level despite the time of the classes. These observations provide directions for further research in maximizing student engagement by a more reasonable arrangement of class schedule according to the nature of each course.
    
    \item \textbf{Engagement and thermal comfort}. In the background survey, most students agree that \textit{ 'When I am engaged in class, I could get distracted when the room is too hot or too cold'} (see Section ~\ref{sec: indoor environmental data}). As another investigative point, Figure ~\ref{fig: thermal comfort} shows the relationship between self-reported engagement and thermal preference (i.e., warmer, cooler, no change) \cite{de2002thermal} of the students in class. The results show that students who feel the room thermally comfortable have a higher overall engagement level compared to other groups. In particular, students who prefer a cooler environment usually have the lowest cognition engagement. This reminds us that creating the right thermally comfortable environment is necessary to improve student engagement in class, especially considering the individual differences in thermal sensation \cite{farhan2015predicting}.
    
    \item \textbf{Real-time measurement}. Students' engagement level during a class may vary with the learning content and teaching style. Real-time anonymous engagement tracking can provide teachers with student engagement level and help teachers understand the impact of different teaching contents on student engagement, thereby better adjusting teaching speed and teaching methods. However, the challenge is how to obtain the fine-grained ground truth of student engagement multiple times during the class without disturbing students' studying. One potential approach is ecological momentary assessment (EMA) \cite{king2019micro} which repeatedly prompts students to report their engagement level. Though EMA is usually considered a good method of \textit{in situ} data collection, if students need to answer EMA, they may be disturbed and distracted in class. Overall, ground truth data collection is challenging and more reliable methods need to be investigated.
\end{itemize}

\section{Implications and Limitations}

\label{sec:implications and limitations}
\textcolor{black}{This research addresses the possibility of automatically predicting students' in-class emotional, behavioural and cognitive engagement using wearable and indoor sensing technology, which provides opportunities for the future design of feedback systems in the classroom. The feedback system has the potential to benefit both teachers and students.}

\textcolor{black}{Teacher plays an important role in influencing student engagement \cite{fredricks2004school}. With the feedback from students after each class, teachers can evaluate, and, if necessary, adapt or change teaching strategies (e.g., increase time for student thinking, allow students time to write, assign reporters for small groups \cite{tanner2013structure}) for creating the right learning climate to keep students engaged \cite{hattie2007power}. For instance, when teachers focus more on academics and fail to create a positive social learning environment, students are likely to be emotionally disengaged and worried about making mistakes. Contrarily, when teachers focus more on the social dimension and neglect the intellectual dimension, students possibly experience low cognitive engagement for learning \cite{stipek2002good,fredricks2004school}. With such a feedback system, teachers can observe multidimensional student engagement and create the intellectually challenging and socially supportive learning environment.}

\textcolor{black}{Further, if this system is deployed, using \textit{n-Gage}, teachers can take timely measures to improve learning experience for students, such as planning learning schedules, re-engaging students with the low engagement, and ventilating the room to let the fresh air in. While overcoming student disengagement is complicated, we do believe teachers can benefit from the engagement feedback of students after every class instead of few times in a term \cite{cantrell2013ensuring,di2018unobtrusive}, contributing to higher student achievements and protecting students from dropping out of school \cite{fredricks2004school}.}


\textcolor{black}{Students wearing wristbands are able to self-track their multidimensional in-class engagement, which positively influences academic achievements and is usually regarded as the predictor of learning outcomes \cite{connell1994educational,marks2000student,fredricks2004school}. Being conscious of in-class engagement is an effective \textit{quantified-self}  \cite{rivera2012applying,eynon2015quantified} approach to promote self-regulation and reflective learning \cite{black2009developing} for students. Once students are aware of how much effort they are putting into learning, they can work towards their personal goals by optimizing their study practices and learning strategies (e.g., practice active listening and thinking, make study plans for different subjects) \cite{eynon2015quantified,arnold2017student}. Additional strategies such as gamification \cite{cronk2012using,buckley2016gamification} can also be deployed along with \textit{
n-Gage} measurements. }



\textcolor{black}{For real-world deployments, the feedback system can still work when only a subset of sensors available (see Section \ref{sec: results sensors}). For instance, when there are no indoor sensors installed, wearable sensors can be used for accurate engagement prediction especially for the emotional engagement. The system can also allow more sensors to be integrated in the future when becomes available. }

The current studies have some limitations that needed to be addressed in future research. Firstly, collecting data from more student participants in the same class may bring new opportunities for data analysis. There are 59 Year 10 students in total, but only 23 students voluntarily become participants and wear wristbands. Compared to students who did not participate,  participants may share some similar personality traits and have higher potential to engage in class most 
of the time. 

Secondly, we agree that collecting the ground truth of student engagement is challenging because we need to find a compromise between taking long psychological surveys for more accurate measurement and enabling students to complete surveys faster without affecting their study or rest. Therefore, a more robust way of evaluating multidimensional student engagement needs to be investigated in the future.

\begin{figure}
    \centering
    \includegraphics[width=0.7\textwidth]{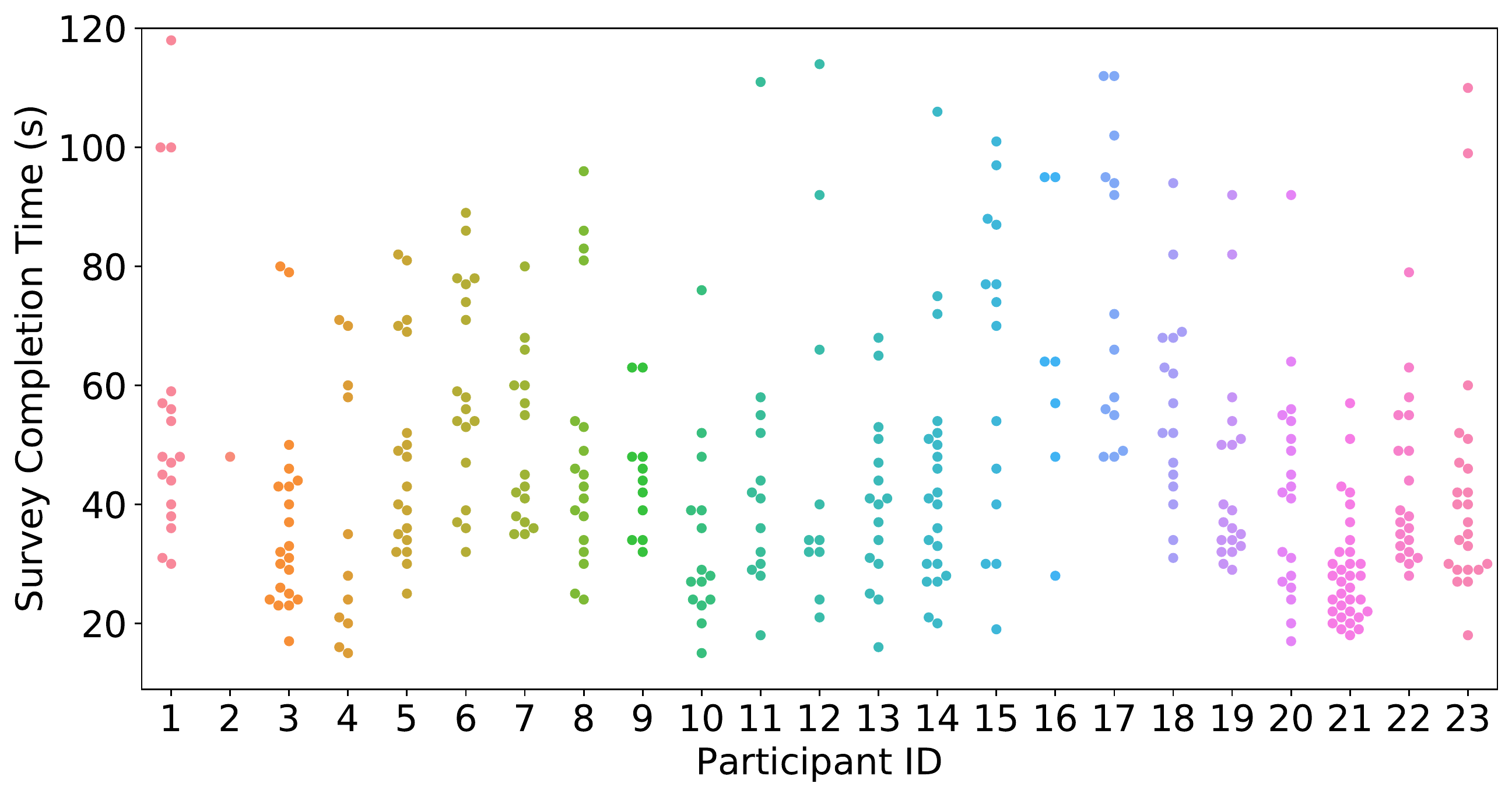}
    \caption{Survey completion time for different participants. Each point represents the survey completion time for one response.}
    \label{fig:completion time}
\end{figure}

Thirdly, the quality of survey responses varies. Online surveys are conducted 3 times a day, and the total response rate is 35.3\%. Since completing surveys multiple times a day may become a burden, students are likely to answer the questions unseriously. Therefore, in this study, we only encourage rather than urge them to complete the survey, which to a certain extent guarantees the quality of responses. Figure \ref{fig:completion time} shows the survey completion time for all responses from participants. Most participants complete the survey in 30 to 50 seconds, but some participants complete the survey in less than 15 seconds. Though the survey completion time may be affected by many factors and varies from person to person, it is still one of the indicators of response quality \cite{malhotra2008completion}. In future research, it will be interesting to explore patterns from survey completion time data and assign appropriate weights to survey response for more accurate prediction of student engagement. 

During the in-situ data collection, the data recorded by the wristband is not always continuous. For many reasons, we face a considerable loss of data: (1) on each school day, an average of 2 to 4 participants got sick leave and cannot wear a wristband; (2) 6 participants went abroad to a study program on the second week of data collection; (3) students were curious about the wristbands, especially in the first few days, and they pressed the button again and again out of curiosity. Some students accidentally closed their wristbands, so their data was lost for hours or even the whole day. 

Though significant efforts have been made to make the maximal use of the collected data, 32.17\% traces must to be removed from the analysis due to the loss of survey data, the incomplete data during the class, the presence of long-time of flat responses, artifacts and quantization errors as discussed in Section \ref{sec: preprocessing}. Despite the fact that we have cleaned and pre-processed wearable data to eliminate noises, collecting physiological data in the wild still faces huge challenges, especially for young students. In our research, one of the main noise is from the poor contact between the sensors and skin, which can be fixed by tightening the wrist strap to the skin. However, this will also increase awareness of wristband during class, resulting in student in-class disengagement and even more motion artifacts. 

\section{Conclusion and Future Work}
\label{sec:conclusion}

In this research, we propose \textit{n-Gage}, an engagement sensing system that can capture students' physiological responses, physical movements, and environmental changes to infer multidimensional engagement (behavioural, emotional and cognitive engagement) level in class. We evaluate the system by combining weather station data and wearable data collected from 23 Year 10 students and 6 teachers over 144 classes in 4 weeks in a high school. Some new features are proposed to characterize different aspects of student engagement. Extensive experiment results show that \textit{n-Gage} can predict student behavioural, emotional and cognitive engagement score (1 is the lowest score and 5 is the highest score) with an average MAE of 0.788 and RMSE of 0.975. We further demonstrate the most influential features and how different sensor combinations/school subjects affect student engagement. Finally, we show some interesting findings that the maximal CO$_2$ level is highly negatively correlated with student cognitive engagement; class time (morning, noon and afternoon) and thermal preference (warmer, cooler or no change) may affect the level of student engagement, which provides beneficial insights for educators and school managers to improve student learning engagement in high school.

Though not perfect, we believe that \textit{n-Gage} is still a very promising first-step towards multidimensional in-class engagement tracking for students. As a contribution, \textcolor{black}{\textit{n-Gage} can indicate the future design of feedback system, assisting students and teachers in a variety of ways (e.g., promoting students' self-regulation and reflective learning, helping teachers create a right learning climate for students). 
In the future, we plan to involve more participants of different ages from different schools in data collection.} The expansion of the dataset will help us get better and more precise results. Also, we hope to investigate more factors that may affect students' multi-engagement, such as personality, mood and behavioural habits.

\section*{Acknowledgement}
This research is supported by the Australian Government through the Australian Research Council's Linkage Projects funding scheme (project LP150100246). This paper is also a contribution to the IEA EBC Annex 79.

\bibliographystyle{ACM-Reference-Format}
\bibliography{nanref}


\begin{thebibliography}{96}


\ifx \showCODEN    \undefined \def \showCODEN     #1{\unskip}     \fi
\ifx \showDOI      \undefined \def \showDOI       #1{#1}\fi
\ifx \showISBNx    \undefined \def \showISBNx     #1{\unskip}     \fi
\ifx \showISBNxiii \undefined \def \showISBNxiii  #1{\unskip}     \fi
\ifx \showISSN     \undefined \def \showISSN      #1{\unskip}     \fi
\ifx \showLCCN     \undefined \def \showLCCN      #1{\unskip}     \fi
\ifx \shownote     \undefined \def \shownote      #1{#1}          \fi
\ifx \showarticletitle \undefined \def \showarticletitle #1{#1}   \fi
\ifx \showURL      \undefined \def \showURL       {\relax}        \fi
\providecommand\bibfield[2]{#2}
\providecommand\bibinfo[2]{#2}
\providecommand\natexlab[1]{#1}
\providecommand\showeprint[2][]{arXiv:#2}

\bibitem[\protect\citeauthoryear{Ahuja, Kim, Xhakaj, Varga, Xie, Zhang,
  Townsend, Harrison, Ogan, and Agarwal}{Ahuja et~al\mbox{.}}{2019}]%
        {ahuja2019edusense}
\bibfield{author}{\bibinfo{person}{Karan Ahuja}, \bibinfo{person}{Dohyun Kim},
  \bibinfo{person}{Franceska Xhakaj}, \bibinfo{person}{Virag Varga},
  \bibinfo{person}{Anne Xie}, \bibinfo{person}{Stanley Zhang},
  \bibinfo{person}{Jay~Eric Townsend}, \bibinfo{person}{Chris Harrison},
  \bibinfo{person}{Amy Ogan}, {and} \bibinfo{person}{Yuvraj Agarwal}.}
  \bibinfo{year}{2019}\natexlab{}.
\newblock \showarticletitle{EduSense: Practical classroom sensing at Scale}.
\newblock \bibinfo{journal}{\emph{Proceedings of the ACM on Interactive,
  Mobile, Wearable and Ubiquitous Technologies}} \bibinfo{volume}{3},
  \bibinfo{number}{3} (\bibinfo{year}{2019}), \bibinfo{pages}{71}.
\newblock


\bibitem[\protect\citeauthoryear{Ahuja, Agarwal, Mahajan, Mehta, and
  Kapadia}{Ahuja et~al\mbox{.}}{2003}]%
        {ahuja2003gsr}
\bibfield{author}{\bibinfo{person}{Nutan~D Ahuja}, \bibinfo{person}{Amit~K
  Agarwal}, \bibinfo{person}{Ninad~M Mahajan}, \bibinfo{person}{Naresh~H
  Mehta}, {and} \bibinfo{person}{Hatim~N Kapadia}.}
  \bibinfo{year}{2003}\natexlab{}.
\newblock \showarticletitle{GSR and HRV: Its Application in Clinical
  Diagnosis}. In \bibinfo{booktitle}{\emph{16th IEEE Symposium Computer-Based
  Medical Systems, 2003. Proceedings.}} IEEE, \bibinfo{pages}{279--283}.
\newblock


\bibitem[\protect\citeauthoryear{Andreassi}{Andreassi}{2010}]%
        {andreassi2010psychophysiology}
\bibfield{author}{\bibinfo{person}{John~L Andreassi}.}
  \bibinfo{year}{2010}\natexlab{}.
\newblock \bibinfo{booktitle}{\emph{Psychophysiology: Human Behavior and
  Physiological Response}}.
\newblock \bibinfo{publisher}{Psychology Press}.
\newblock


\bibitem[\protect\citeauthoryear{Appelhans and Luecken}{Appelhans and
  Luecken}{2006}]%
        {appelhans2006heart}
\bibfield{author}{\bibinfo{person}{Bradley~M Appelhans} {and}
  \bibinfo{person}{Linda~J Luecken}.} \bibinfo{year}{2006}\natexlab{}.
\newblock \showarticletitle{Heart Rate Variability as an Index of Regulated
  Emotional Responding}.
\newblock \bibinfo{journal}{\emph{Review of general psychology}}
  \bibinfo{volume}{10}, \bibinfo{number}{3} (\bibinfo{year}{2006}),
  \bibinfo{pages}{229--240}.
\newblock


\bibitem[\protect\citeauthoryear{Appleton, Christenson, Kim, and
  Reschly}{Appleton et~al\mbox{.}}{2006}]%
        {appleton2006measuring}
\bibfield{author}{\bibinfo{person}{James~J Appleton}, \bibinfo{person}{Sandra~L
  Christenson}, \bibinfo{person}{Dongjin Kim}, {and} \bibinfo{person}{Amy~L
  Reschly}.} \bibinfo{year}{2006}\natexlab{}.
\newblock \showarticletitle{Measuring Cognitive and Psychological Engagement:
  Validation of the Student Engagement Instrument}.
\newblock \bibinfo{journal}{\emph{Journal of school psychology}}
  \bibinfo{volume}{44}, \bibinfo{number}{5} (\bibinfo{year}{2006}),
  \bibinfo{pages}{427--445}.
\newblock


\bibitem[\protect\citeauthoryear{Arnold, Karcher, Wright, and McKay}{Arnold
  et~al\mbox{.}}{2017}]%
        {arnold2017student}
\bibfield{author}{\bibinfo{person}{Kimberly~E Arnold}, \bibinfo{person}{Brandon
  Karcher}, \bibinfo{person}{Casey~V Wright}, {and} \bibinfo{person}{James
  McKay}.} \bibinfo{year}{2017}\natexlab{}.
\newblock \showarticletitle{Student Empowerment, Awareness, and Self-regulation
  Through a Quantified-self Student Tool}. In
  \bibinfo{booktitle}{\emph{Proceedings of the Seventh International Learning
  Analytics \& Knowledge Conference}}. \bibinfo{pages}{526--527}.
\newblock


\bibitem[\protect\citeauthoryear{Bakker, Pechenizkiy, and Sidorova}{Bakker
  et~al\mbox{.}}{2011}]%
        {bakker2011s}
\bibfield{author}{\bibinfo{person}{Jorn Bakker}, \bibinfo{person}{Mykola
  Pechenizkiy}, {and} \bibinfo{person}{Natalia Sidorova}.}
  \bibinfo{year}{2011}\natexlab{}.
\newblock \showarticletitle{What's Your Current Stress Level? Detection of
  Stress Patterns from GSR Sensor Data}. In \bibinfo{booktitle}{\emph{2011 IEEE
  11th international conference on data mining workshops}}. IEEE,
  \bibinfo{pages}{573--580}.
\newblock


\bibitem[\protect\citeauthoryear{Benesty, Chen, Huang, and Cohen}{Benesty
  et~al\mbox{.}}{2009}]%
        {benesty2009pearson}
\bibfield{author}{\bibinfo{person}{Jacob Benesty}, \bibinfo{person}{Jingdong
  Chen}, \bibinfo{person}{Yiteng Huang}, {and} \bibinfo{person}{Israel Cohen}.}
  \bibinfo{year}{2009}\natexlab{}.
\newblock \showarticletitle{Pearson Correlation Coefficient}.
\newblock In \bibinfo{booktitle}{\emph{Noise reduction in speech processing}}.
  \bibinfo{publisher}{Springer}, \bibinfo{pages}{1--4}.
\newblock


\bibitem[\protect\citeauthoryear{Berka, Levendowski, Lumicao, Yau, Davis,
  Zivkovic, Olmstead, Tremoulet, and Craven}{Berka et~al\mbox{.}}{2007}]%
        {berka2007eeg}
\bibfield{author}{\bibinfo{person}{Chris Berka}, \bibinfo{person}{Daniel~J
  Levendowski}, \bibinfo{person}{Michelle~N Lumicao}, \bibinfo{person}{Alan
  Yau}, \bibinfo{person}{Gene Davis}, \bibinfo{person}{Vladimir~T Zivkovic},
  \bibinfo{person}{Richard~E Olmstead}, \bibinfo{person}{Patrice~D Tremoulet},
  {and} \bibinfo{person}{Patrick~L Craven}.} \bibinfo{year}{2007}\natexlab{}.
\newblock \showarticletitle{EEG Correlates of Task Engagement and Mental
  Workload in Vigilance, Learning, and Memory Tasks}.
\newblock \bibinfo{journal}{\emph{Aviation, space, and environmental medicine}}
  \bibinfo{volume}{78}, \bibinfo{number}{5} (\bibinfo{year}{2007}),
  \bibinfo{pages}{B231--B244}.
\newblock


\bibitem[\protect\citeauthoryear{Black and Wiliam}{Black and Wiliam}{2009}]%
        {black2009developing}
\bibfield{author}{\bibinfo{person}{Paul Black} {and} \bibinfo{person}{Dylan
  Wiliam}.} \bibinfo{year}{2009}\natexlab{}.
\newblock \showarticletitle{Developing the Theory of Formative Assessment}.
\newblock \bibinfo{journal}{\emph{Educational Assessment, Evaluation and
  Accountability (formerly: Journal of Personnel Evaluation in Education)}}
  \bibinfo{volume}{21}, \bibinfo{number}{1} (\bibinfo{year}{2009}),
  \bibinfo{pages}{5}.
\newblock


\bibitem[\protect\citeauthoryear{Blanchard, Bixler, Joyce, and
  D'Mello}{Blanchard et~al\mbox{.}}{2014}]%
        {blanchard2014automated}
\bibfield{author}{\bibinfo{person}{Nathaniel Blanchard},
  \bibinfo{person}{Robert Bixler}, \bibinfo{person}{Tera Joyce}, {and}
  \bibinfo{person}{Sidney D'Mello}.} \bibinfo{year}{2014}\natexlab{}.
\newblock \showarticletitle{Automated Physiological-based Detection of Mind
  Wandering During Learning}. In \bibinfo{booktitle}{\emph{International
  Conference on Intelligent Tutoring Systems}}. Springer,
  \bibinfo{pages}{55--60}.
\newblock


\bibitem[\protect\citeauthoryear{Boucsein}{Boucsein}{2012}]%
        {boucsein2012electrodermal}
\bibfield{author}{\bibinfo{person}{Wolfram Boucsein}.}
  \bibinfo{year}{2012}\natexlab{}.
\newblock \showarticletitle{Electrodermal Activity: Springer Science \&
  Business Media}.
\newblock \bibinfo{journal}{\emph{Broek, EL vd, Schut, MH, Westerink, JHDM,
  Herk, J. v., \& Tuinenbreijer, K}} (\bibinfo{year}{2012}).
\newblock


\bibitem[\protect\citeauthoryear{Buckley and Doyle}{Buckley and Doyle}{2016}]%
        {buckley2016gamification}
\bibfield{author}{\bibinfo{person}{Patrick Buckley} {and}
  \bibinfo{person}{Elaine Doyle}.} \bibinfo{year}{2016}\natexlab{}.
\newblock \showarticletitle{Gamification and student motivation}.
\newblock \bibinfo{journal}{\emph{Interactive learning environments}}
  \bibinfo{volume}{24}, \bibinfo{number}{6} (\bibinfo{year}{2016}),
  \bibinfo{pages}{1162--1175}.
\newblock


\bibitem[\protect\citeauthoryear{Cacioppo, Tassinary, and Berntson}{Cacioppo
  et~al\mbox{.}}{2007}]%
        {cacioppo2007handbook}
\bibfield{author}{\bibinfo{person}{John~T Cacioppo}, \bibinfo{person}{Louis~G
  Tassinary}, {and} \bibinfo{person}{Gary Berntson}.}
  \bibinfo{year}{2007}\natexlab{}.
\newblock \bibinfo{booktitle}{\emph{Handbook of Psychophysiology}}.
\newblock \bibinfo{publisher}{Cambridge University Press}.
\newblock


\bibitem[\protect\citeauthoryear{Camm, Malik, Bigger, Breithardt, Cerutti,
  Cohen, Coumel, Fallen, Kennedy, Kleiger, et~al\mbox{.}}{Camm
  et~al\mbox{.}}{1996}]%
        {camm1996heart}
\bibfield{author}{\bibinfo{person}{A~John Camm}, \bibinfo{person}{Marek Malik},
  \bibinfo{person}{J~Thomas Bigger}, \bibinfo{person}{G{\"u}nter Breithardt},
  \bibinfo{person}{Sergio Cerutti}, \bibinfo{person}{Richard~J Cohen},
  \bibinfo{person}{Philippe Coumel}, \bibinfo{person}{Ernest~L Fallen},
  \bibinfo{person}{Harold~L Kennedy}, \bibinfo{person}{RE Kleiger},
  {et~al\mbox{.}}} \bibinfo{year}{1996}\natexlab{}.
\newblock \showarticletitle{Heart Rate Variability: Standards of Measurement,
  physiological interpretation and clinical use. Task Force of the European
  Society of Cardiology and the North American Society of Pacing and
  Electrophysiology}.
\newblock  (\bibinfo{year}{1996}).
\newblock


\bibitem[\protect\citeauthoryear{Canento, Fred, Silva, Gamboa, and
  Louren{\c{c}}o}{Canento et~al\mbox{.}}{2011}]%
        {canento2011multimodal}
\bibfield{author}{\bibinfo{person}{Filipe Canento}, \bibinfo{person}{Ana Fred},
  \bibinfo{person}{Hugo Silva}, \bibinfo{person}{Hugo Gamboa}, {and}
  \bibinfo{person}{Andr{\'e} Louren{\c{c}}o}.} \bibinfo{year}{2011}\natexlab{}.
\newblock \showarticletitle{Multimodal Biosignal Sensor Data Handling for
  Emotion Recognition}. In \bibinfo{booktitle}{\emph{SENSORS, 2011 IEEE}}.
  IEEE, \bibinfo{pages}{647--650}.
\newblock


\bibitem[\protect\citeauthoryear{Cantrell and Kane}{Cantrell and Kane}{2013}]%
        {cantrell2013ensuring}
\bibfield{author}{\bibinfo{person}{Steven Cantrell} {and}
  \bibinfo{person}{Thomas~J Kane}.} \bibinfo{year}{2013}\natexlab{}.
\newblock \showarticletitle{Ensuring Fair and Reliable Measures of Effective
  Teaching: Culminating Findings from the MET Project's Three-year Study}.
\newblock \bibinfo{journal}{\emph{MET Project Research Paper}}
  (\bibinfo{year}{2013}).
\newblock


\bibitem[\protect\citeauthoryear{Chai and Draxler}{Chai and Draxler}{2014}]%
        {chai2014root}
\bibfield{author}{\bibinfo{person}{Tianfeng Chai} {and}
  \bibinfo{person}{Roland~R Draxler}.} \bibinfo{year}{2014}\natexlab{}.
\newblock \showarticletitle{Root Mean Square Error (RMSE) or Mean Absolute
  Error (MAE)?--Arguments Against Avoiding RMSE in the Literature}.
\newblock \bibinfo{journal}{\emph{Geoscientific model development}}
  \bibinfo{volume}{7}, \bibinfo{number}{3} (\bibinfo{year}{2014}),
  \bibinfo{pages}{1247--1250}.
\newblock


\bibitem[\protect\citeauthoryear{Choi, Kim, Kwon, Kim, Ryu, and Park}{Choi
  et~al\mbox{.}}{2017}]%
        {choi2017heart}
\bibfield{author}{\bibinfo{person}{Kwang-Ho Choi}, \bibinfo{person}{Junbeom
  Kim}, \bibinfo{person}{O~Sang Kwon}, \bibinfo{person}{Min~Ji Kim},
  \bibinfo{person}{Yeon~Hee Ryu}, {and} \bibinfo{person}{Ji-Eun Park}.}
  \bibinfo{year}{2017}\natexlab{}.
\newblock \showarticletitle{Is Heart Rate Variability (HRV) an Adequate Tool
  for Evaluating Human Emotions?--A Focus on the Use of the International
  Affective Picture System (IAPS)}.
\newblock \bibinfo{journal}{\emph{Psychiatry research}}  \bibinfo{volume}{251}
  (\bibinfo{year}{2017}), \bibinfo{pages}{192--196}.
\newblock


\bibitem[\protect\citeauthoryear{Connell, Spencer, and Aber}{Connell
  et~al\mbox{.}}{1994}]%
        {connell1994educational}
\bibfield{author}{\bibinfo{person}{James~Patrick Connell},
  \bibinfo{person}{Margaret~Beale Spencer}, {and} \bibinfo{person}{J~Lawrence
  Aber}.} \bibinfo{year}{1994}\natexlab{}.
\newblock \showarticletitle{Educational Risk and Resilience in African-American
  Youth: Context, Self, Action, and Outcomes in School}.
\newblock \bibinfo{journal}{\emph{Child development}} \bibinfo{volume}{65},
  \bibinfo{number}{2} (\bibinfo{year}{1994}), \bibinfo{pages}{493--506}.
\newblock


\bibitem[\protect\citeauthoryear{Corno and Mandinach}{Corno and
  Mandinach}{1983}]%
        {corno1983role}
\bibfield{author}{\bibinfo{person}{Lyn Corno} {and} \bibinfo{person}{Ellen~B
  Mandinach}.} \bibinfo{year}{1983}\natexlab{}.
\newblock \showarticletitle{The Role of Cognitive Engagement in Classroom
  Learning and Motivation}.
\newblock \bibinfo{journal}{\emph{Educational psychologist}}
  \bibinfo{volume}{18}, \bibinfo{number}{2} (\bibinfo{year}{1983}),
  \bibinfo{pages}{88--108}.
\newblock


\bibitem[\protect\citeauthoryear{Council et~al\mbox{.}}{Council
  et~al\mbox{.}}{2003}]%
        {national2003engaging}
\bibfield{author}{\bibinfo{person}{National~Research Council} {et~al\mbox{.}}}
  \bibinfo{year}{2003}\natexlab{}.
\newblock \bibinfo{booktitle}{\emph{Engaging Schools: Fostering High School
  Students' Motivation to Learn}}.
\newblock \bibinfo{publisher}{National Academies Press}.
\newblock


\bibitem[\protect\citeauthoryear{Critchley}{Critchley}{2002}]%
        {critchley2002electrodermal}
\bibfield{author}{\bibinfo{person}{Hugo~D Critchley}.}
  \bibinfo{year}{2002}\natexlab{}.
\newblock \showarticletitle{Electrodermal Responses: What Happens in the
  Brain}.
\newblock \bibinfo{journal}{\emph{The Neuroscientist}} \bibinfo{volume}{8},
  \bibinfo{number}{2} (\bibinfo{year}{2002}), \bibinfo{pages}{132--142}.
\newblock


\bibitem[\protect\citeauthoryear{Cronk}{Cronk}{2012}]%
        {cronk2012using}
\bibfield{author}{\bibinfo{person}{Marguerite Cronk}.}
  \bibinfo{year}{2012}\natexlab{}.
\newblock \showarticletitle{Using Gamification to Increase Student Engagement
  and Participation in Class Discussion}. In \bibinfo{booktitle}{\emph{EdMedia+
  Innovate Learning}}. Association for the Advancement of Computing in
  Education (AACE), \bibinfo{pages}{311--315}.
\newblock


\bibitem[\protect\citeauthoryear{De~Dear and Brager}{De~Dear and
  Brager}{2002}]%
        {de2002thermal}
\bibfield{author}{\bibinfo{person}{Richard~J De~Dear} {and}
  \bibinfo{person}{Gail~S Brager}.} \bibinfo{year}{2002}\natexlab{}.
\newblock \showarticletitle{Thermal Comfort in Naturally Ventilated Buildings:
  Revisions to ASHRAE Standard 55}.
\newblock \bibinfo{journal}{\emph{Energy and buildings}} \bibinfo{volume}{34},
  \bibinfo{number}{6} (\bibinfo{year}{2002}), \bibinfo{pages}{549--561}.
\newblock


\bibitem[\protect\citeauthoryear{Deldari, Liono, Salim, and Smith}{Deldari
  et~al\mbox{.}}{2019}]%
        {deldari2016inferring}
\bibfield{author}{\bibinfo{person}{Shohreh Deldari}, \bibinfo{person}{Jonathan
  Liono}, \bibinfo{person}{Flora~D Salim}, {and} \bibinfo{person}{Daniel~V
  Smith}.} \bibinfo{year}{2019}\natexlab{}.
\newblock \showarticletitle{Inferring Work Routines and Behavior Deviations
  with Life-logging Sensor Data}.
\newblock  (\bibinfo{year}{2019}).
\newblock


\bibitem[\protect\citeauthoryear{Di~Lascio, Gashi, and Santini}{Di~Lascio
  et~al\mbox{.}}{2018}]%
        {di2018unobtrusive}
\bibfield{author}{\bibinfo{person}{Elena Di~Lascio}, \bibinfo{person}{Shkurta
  Gashi}, {and} \bibinfo{person}{Silvia Santini}.}
  \bibinfo{year}{2018}\natexlab{}.
\newblock \showarticletitle{Unobtrusive Assessment of Students' Emotional
  Engagement During Lectures Using Electrodermal Activity Sensors}.
\newblock \bibinfo{journal}{\emph{Proceedings of the ACM on Interactive,
  Mobile, Wearable and Ubiquitous Technologies}} \bibinfo{volume}{2},
  \bibinfo{number}{3} (\bibinfo{year}{2018}), \bibinfo{pages}{103}.
\newblock


\bibitem[\protect\citeauthoryear{Elith, Leathwick, and Hastie}{Elith
  et~al\mbox{.}}{2008}]%
        {elith2008working}
\bibfield{author}{\bibinfo{person}{Jane Elith}, \bibinfo{person}{John~R
  Leathwick}, {and} \bibinfo{person}{Trevor Hastie}.}
  \bibinfo{year}{2008}\natexlab{}.
\newblock \showarticletitle{A Working Guide to Boosted Regression Trees}.
\newblock \bibinfo{journal}{\emph{Journal of Animal Ecology}}
  \bibinfo{volume}{77}, \bibinfo{number}{4} (\bibinfo{year}{2008}),
  \bibinfo{pages}{802--813}.
\newblock


\bibitem[\protect\citeauthoryear{Eynon}{Eynon}{2015}]%
        {eynon2015quantified}
\bibfield{author}{\bibinfo{person}{Rebecca Eynon}.}
  \bibinfo{year}{2015}\natexlab{}.
\newblock \bibinfo{title}{The Quantified Self for Learning: Critical Questions
  for Education}.
\newblock
\newblock


\bibitem[\protect\citeauthoryear{Farhan, Pattipati, Wang, and Luh}{Farhan
  et~al\mbox{.}}{2015}]%
        {farhan2015predicting}
\bibfield{author}{\bibinfo{person}{Asma~Ahmad Farhan}, \bibinfo{person}{Krishna
  Pattipati}, \bibinfo{person}{Bing Wang}, {and} \bibinfo{person}{Peter Luh}.}
  \bibinfo{year}{2015}\natexlab{}.
\newblock \showarticletitle{Predicting Individual Thermal Comfort Using Machine
  Learning Algorithms}. In \bibinfo{booktitle}{\emph{2015 IEEE International
  Conference on Automation Science and Engineering (CASE)}}. IEEE,
  \bibinfo{pages}{708--713}.
\newblock


\bibitem[\protect\citeauthoryear{Finn}{Finn}{1989}]%
        {finn1989withdrawing}
\bibfield{author}{\bibinfo{person}{Jeremy~D Finn}.}
  \bibinfo{year}{1989}\natexlab{}.
\newblock \showarticletitle{Withdrawing from School}.
\newblock \bibinfo{journal}{\emph{Review of educational research}}
  \bibinfo{volume}{59}, \bibinfo{number}{2} (\bibinfo{year}{1989}),
  \bibinfo{pages}{117--142}.
\newblock


\bibitem[\protect\citeauthoryear{Finn, Pannozzo, and Voelkl}{Finn
  et~al\mbox{.}}{1995}]%
        {finn1995disruptive}
\bibfield{author}{\bibinfo{person}{Jeremy~D Finn}, \bibinfo{person}{Gina~M
  Pannozzo}, {and} \bibinfo{person}{Kristin~E Voelkl}.}
  \bibinfo{year}{1995}\natexlab{}.
\newblock \showarticletitle{Disruptive and Inattentive-withdrawn Behavior and
  Achievement Among Fourth Graders}.
\newblock \bibinfo{journal}{\emph{The Elementary School Journal}}
  \bibinfo{volume}{95}, \bibinfo{number}{5} (\bibinfo{year}{1995}),
  \bibinfo{pages}{421--434}.
\newblock


\bibitem[\protect\citeauthoryear{Finn and Rock}{Finn and Rock}{1997}]%
        {finn1997academic}
\bibfield{author}{\bibinfo{person}{Jeremy~D Finn} {and}
  \bibinfo{person}{Donald~A Rock}.} \bibinfo{year}{1997}\natexlab{}.
\newblock \showarticletitle{Academic Success Among Students at Risk for School
  Failure}.
\newblock \bibinfo{journal}{\emph{Journal of applied psychology}}
  \bibinfo{volume}{82}, \bibinfo{number}{2} (\bibinfo{year}{1997}),
  \bibinfo{pages}{221}.
\newblock


\bibitem[\protect\citeauthoryear{Fredricks, Blumenfeld, and Paris}{Fredricks
  et~al\mbox{.}}{2004}]%
        {fredricks2004school}
\bibfield{author}{\bibinfo{person}{Jennifer~A Fredricks},
  \bibinfo{person}{Phyllis~C Blumenfeld}, {and} \bibinfo{person}{Alison~H
  Paris}.} \bibinfo{year}{2004}\natexlab{}.
\newblock \showarticletitle{School Engagement: Potential of the Concept, State
  of the Evidence}.
\newblock \bibinfo{journal}{\emph{Review of educational research}}
  \bibinfo{volume}{74}, \bibinfo{number}{1} (\bibinfo{year}{2004}),
  \bibinfo{pages}{59--109}.
\newblock


\bibitem[\protect\citeauthoryear{Fredricks and McColskey}{Fredricks and
  McColskey}{2012}]%
        {fredricks2012measurement}
\bibfield{author}{\bibinfo{person}{Jennifer~A Fredricks} {and}
  \bibinfo{person}{Wendy McColskey}.} \bibinfo{year}{2012}\natexlab{}.
\newblock \showarticletitle{The Measurement of Student Engagement: A
  Comparative Analysis of Various Methods and Student Self-report Instruments}.
\newblock In \bibinfo{booktitle}{\emph{Handbook of research on student
  engagement}}. \bibinfo{publisher}{Springer}, \bibinfo{pages}{763--782}.
\newblock


\bibitem[\protect\citeauthoryear{Friedman, Hastie, and Tibshirani}{Friedman
  et~al\mbox{.}}{2001}]%
        {friedman2001elements}
\bibfield{author}{\bibinfo{person}{Jerome Friedman}, \bibinfo{person}{Trevor
  Hastie}, {and} \bibinfo{person}{Robert Tibshirani}.}
  \bibinfo{year}{2001}\natexlab{}.
\newblock \bibinfo{booktitle}{\emph{The Elements of Statistical Learning}}.
  Vol.~\bibinfo{volume}{1}.
\newblock \bibinfo{publisher}{Springer series in statistics New York}.
\newblock


\bibitem[\protect\citeauthoryear{Fuller, Karunaratne, Naidu, Exintaris, Short,
  Wolcott, Singleton, and White}{Fuller et~al\mbox{.}}{2018}]%
        {fuller2018development}
\bibfield{author}{\bibinfo{person}{Kathryn~A Fuller},
  \bibinfo{person}{Nilushi~S Karunaratne}, \bibinfo{person}{Som Naidu},
  \bibinfo{person}{Betty Exintaris}, \bibinfo{person}{Jennifer~L Short},
  \bibinfo{person}{Michael~D Wolcott}, \bibinfo{person}{Scott Singleton}, {and}
  \bibinfo{person}{Paul~J White}.} \bibinfo{year}{2018}\natexlab{}.
\newblock \showarticletitle{Development of a self-report Instrument for
  Measuring In-class Student Engagement Reveals that Pretending to Engage is a
  Significant Unrecognized Problem}.
\newblock \bibinfo{journal}{\emph{PloS one}} \bibinfo{volume}{13},
  \bibinfo{number}{10} (\bibinfo{year}{2018}), \bibinfo{pages}{e0205828}.
\newblock


\bibitem[\protect\citeauthoryear{Gao, Shao, Rahaman, Zhai, David, and
  Salim}{Gao et~al\mbox{.}}{2020}]%
        {gao2020transfer}
\bibfield{author}{\bibinfo{person}{Nan Gao}, \bibinfo{person}{Wei Shao},
  \bibinfo{person}{Mohammad~Saiedur Rahaman}, \bibinfo{person}{Jun Zhai},
  \bibinfo{person}{Klaus David}, {and} \bibinfo{person}{Flora~D Salim}.}
  \bibinfo{year}{2020}\natexlab{}.
\newblock \showarticletitle{Transfer Learning for Thermal Comfort Prediction in
  Multiple Cities}.
\newblock \bibinfo{journal}{\emph{arXiv preprint arXiv:2004.14382}}
  (\bibinfo{year}{2020}).
\newblock


\bibitem[\protect\citeauthoryear{Gao, Shao, and Salim}{Gao
  et~al\mbox{.}}{2019}]%
        {gao2019predicting}
\bibfield{author}{\bibinfo{person}{Nan Gao}, \bibinfo{person}{Wei Shao}, {and}
  \bibinfo{person}{Flora~D Salim}.} \bibinfo{year}{2019}\natexlab{}.
\newblock \showarticletitle{Predicting Personality Traits From Physical
  Activity Intensity}.
\newblock \bibinfo{journal}{\emph{Computer}} \bibinfo{volume}{52},
  \bibinfo{number}{7} (\bibinfo{year}{2019}), \bibinfo{pages}{47--56}.
\newblock


\bibitem[\protect\citeauthoryear{Garbarino, Lai, Bender, Picard, and
  Tognetti}{Garbarino et~al\mbox{.}}{2014}]%
        {garbarino2014empatica}
\bibfield{author}{\bibinfo{person}{Maurizio Garbarino}, \bibinfo{person}{Matteo
  Lai}, \bibinfo{person}{Dan Bender}, \bibinfo{person}{Rosalind~W Picard},
  {and} \bibinfo{person}{Simone Tognetti}.} \bibinfo{year}{2014}\natexlab{}.
\newblock \showarticletitle{Empatica E3—A wearable Wireless Multi-sensor
  Device for Real-time Computerized Biofeedback and Data Acquisition}. In
  \bibinfo{booktitle}{\emph{2014 4th International Conference on Wireless
  Mobile Communication and Healthcare-Transforming Healthcare Through
  Innovations in Mobile and Wireless Technologies (MOBIHEALTH)}}. IEEE,
  \bibinfo{pages}{39--42}.
\newblock


\bibitem[\protect\citeauthoryear{Gashi, Di~Lascio, and Santini}{Gashi
  et~al\mbox{.}}{2019}]%
        {gashi2019using}
\bibfield{author}{\bibinfo{person}{Shkurta Gashi}, \bibinfo{person}{Elena
  Di~Lascio}, {and} \bibinfo{person}{Silvia Santini}.}
  \bibinfo{year}{2019}\natexlab{}.
\newblock \showarticletitle{Using Unobtrusive Wearable Sensors to Measure the
  Physiological Synchrony Between Presenters and Audience Members}.
\newblock \bibinfo{journal}{\emph{Proceedings of the ACM on Interactive,
  Mobile, Wearable and Ubiquitous Technologies}} \bibinfo{volume}{3},
  \bibinfo{number}{1} (\bibinfo{year}{2019}), \bibinfo{pages}{13}.
\newblock


\bibitem[\protect\citeauthoryear{Greco, Valenza, Lanata, Scilingo, and
  Citi}{Greco et~al\mbox{.}}{2015}]%
        {greco2015cvxeda}
\bibfield{author}{\bibinfo{person}{Alberto Greco}, \bibinfo{person}{Gaetano
  Valenza}, \bibinfo{person}{Antonio Lanata}, \bibinfo{person}{Enzo~Pasquale
  Scilingo}, {and} \bibinfo{person}{Luca Citi}.}
  \bibinfo{year}{2015}\natexlab{}.
\newblock \showarticletitle{cvxEDA: A Convex Optimization Approach to
  Electrodermal Activity Processing}.
\newblock \bibinfo{journal}{\emph{IEEE Transactions on Biomedical Engineering}}
  \bibinfo{volume}{63}, \bibinfo{number}{4} (\bibinfo{year}{2015}),
  \bibinfo{pages}{797--804}.
\newblock


\bibitem[\protect\citeauthoryear{Groccia}{Groccia}{2018}]%
        {groccia2018student}
\bibfield{author}{\bibinfo{person}{James~E Groccia}.}
  \bibinfo{year}{2018}\natexlab{}.
\newblock \showarticletitle{What is student engagement?}
\newblock \bibinfo{journal}{\emph{New Directions for Teaching and Learning}}
  \bibinfo{volume}{2018}, \bibinfo{number}{154} (\bibinfo{year}{2018}),
  \bibinfo{pages}{11--20}.
\newblock


\bibitem[\protect\citeauthoryear{Hattie and Timperley}{Hattie and
  Timperley}{2007}]%
        {hattie2007power}
\bibfield{author}{\bibinfo{person}{John Hattie} {and} \bibinfo{person}{Helen
  Timperley}.} \bibinfo{year}{2007}\natexlab{}.
\newblock \showarticletitle{The Power of Feedback}.
\newblock \bibinfo{journal}{\emph{Review of educational research}}
  \bibinfo{volume}{77}, \bibinfo{number}{1} (\bibinfo{year}{2007}),
  \bibinfo{pages}{81--112}.
\newblock


\bibitem[\protect\citeauthoryear{Haverinen-Shaughnessy, Moschandreas, and
  Shaughnessy}{Haverinen-Shaughnessy et~al\mbox{.}}{2011}]%
        {haverinen2011association}
\bibfield{author}{\bibinfo{person}{Ulla Haverinen-Shaughnessy},
  \bibinfo{person}{DJ Moschandreas}, {and} \bibinfo{person}{RJ Shaughnessy}.}
  \bibinfo{year}{2011}\natexlab{}.
\newblock \showarticletitle{Association Between Substandard Classroom
  Ventilation Rates and Students' Academic Achievement}.
\newblock \bibinfo{journal}{\emph{Indoor air}} \bibinfo{volume}{21},
  \bibinfo{number}{2} (\bibinfo{year}{2011}), \bibinfo{pages}{121--131}.
\newblock


\bibitem[\protect\citeauthoryear{Herborn, Graves, Jerem, Evans, Nager,
  McCafferty, and McKeegan}{Herborn et~al\mbox{.}}{2015}]%
        {herborn2015skin}
\bibfield{author}{\bibinfo{person}{Katherine~A Herborn},
  \bibinfo{person}{James~L Graves}, \bibinfo{person}{Paul Jerem},
  \bibinfo{person}{Neil~P Evans}, \bibinfo{person}{Ruedi Nager},
  \bibinfo{person}{Dominic~J McCafferty}, {and} \bibinfo{person}{Dorothy~EF
  McKeegan}.} \bibinfo{year}{2015}\natexlab{}.
\newblock \showarticletitle{Skin Temperature Reveals the Intensity of Acute
  Stress}.
\newblock \bibinfo{journal}{\emph{Physiology \& behavior}}
  \bibinfo{volume}{152} (\bibinfo{year}{2015}), \bibinfo{pages}{225--230}.
\newblock


\bibitem[\protect\citeauthoryear{Hernandez, Riobo, Rozga, Abowd, and
  Picard}{Hernandez et~al\mbox{.}}{2014}]%
        {hernandez2014using}
\bibfield{author}{\bibinfo{person}{Javier Hernandez}, \bibinfo{person}{Ivan
  Riobo}, \bibinfo{person}{Agata Rozga}, \bibinfo{person}{Gregory~D Abowd},
  {and} \bibinfo{person}{Rosalind~W Picard}.} \bibinfo{year}{2014}\natexlab{}.
\newblock \showarticletitle{Using Electrodermal Activity to Recognize Ease of
  Engagement in Children During Social Interactions}. In
  \bibinfo{booktitle}{\emph{Proceedings of the 2014 ACM International Joint
  Conference on Pervasive and Ubiquitous Computing}}. ACM,
  \bibinfo{pages}{307--317}.
\newblock


\bibitem[\protect\citeauthoryear{Hutt, Krasich, Mills, Bosch, White, Brockmole,
  and D'Mello}{Hutt et~al\mbox{.}}{2019}]%
        {hutt2019automated}
\bibfield{author}{\bibinfo{person}{Stephen Hutt}, \bibinfo{person}{Kristina
  Krasich}, \bibinfo{person}{Caitlin Mills}, \bibinfo{person}{Nigel Bosch},
  \bibinfo{person}{Shelby White}, \bibinfo{person}{James~R Brockmole}, {and}
  \bibinfo{person}{Sidney~K D'Mello}.} \bibinfo{year}{2019}\natexlab{}.
\newblock \showarticletitle{Automated Gaze-based Mind Wandering Detection
  During Computerized Learning in Classrooms}.
\newblock \bibinfo{journal}{\emph{User Modeling and User-Adapted Interaction}}
  \bibinfo{volume}{29}, \bibinfo{number}{4} (\bibinfo{year}{2019}),
  \bibinfo{pages}{821--867}.
\newblock


\bibitem[\protect\citeauthoryear{Huynh, Kim, Ko, Balan, and Lee}{Huynh
  et~al\mbox{.}}{2018}]%
        {huynh2018engagemon}
\bibfield{author}{\bibinfo{person}{Sinh Huynh}, \bibinfo{person}{Seungmin Kim},
  \bibinfo{person}{JeongGil Ko}, \bibinfo{person}{Rajesh~Krishna Balan}, {and}
  \bibinfo{person}{Youngki Lee}.} \bibinfo{year}{2018}\natexlab{}.
\newblock \showarticletitle{EngageMon: Multi-Modal Engagement Sensing for
  Mobile Games}.
\newblock \bibinfo{journal}{\emph{Proceedings of the ACM on Interactive,
  Mobile, Wearable and Ubiquitous Technologies}} \bibinfo{volume}{2},
  \bibinfo{number}{1} (\bibinfo{year}{2018}), \bibinfo{pages}{13}.
\newblock


\bibitem[\protect\citeauthoryear{Jiang, Iandoli, Van~Dessel, Liu, and
  Whitehill}{Jiang et~al\mbox{.}}{2019}]%
        {jiang2019measuring}
\bibfield{author}{\bibinfo{person}{Han Jiang}, \bibinfo{person}{Matthew
  Iandoli}, \bibinfo{person}{Steven Van~Dessel}, \bibinfo{person}{Shichao Liu},
  {and} \bibinfo{person}{Jacob Whitehill}.} \bibinfo{year}{2019}\natexlab{}.
\newblock \showarticletitle{Measuring Students' Thermal Comfort and Its Impact
  on Learning}.
\newblock \bibinfo{journal}{\emph{Educational Data Mining}}
  (\bibinfo{year}{2019}).
\newblock


\bibitem[\protect\citeauthoryear{Kane}{Kane}{2020}]%
        {kane}
\bibfield{author}{\bibinfo{person}{Kane}.} \bibinfo{year}{2020}\natexlab{}.
\newblock \bibinfo{title}{What are Safe Levels of CO and CO2 in Rooms?}
\newblock
\newblock
\urldef\tempurl%
\url{https://www.kane.co.uk/knowledge-centre/what-are-safe-levels-of-co-and-co2-in-rooms}
\showURL{%
\tempurl}
\newblock
\shownote{Accessed 2020-07-08.}


\bibitem[\protect\citeauthoryear{Ke, Meng, Finley, Wang, Chen, Ma, Ye, and
  Liu}{Ke et~al\mbox{.}}{2017}]%
        {ke2017lightgbm}
\bibfield{author}{\bibinfo{person}{Guolin Ke}, \bibinfo{person}{Qi Meng},
  \bibinfo{person}{Thomas Finley}, \bibinfo{person}{Taifeng Wang},
  \bibinfo{person}{Wei Chen}, \bibinfo{person}{Weidong Ma},
  \bibinfo{person}{Qiwei Ye}, {and} \bibinfo{person}{Tie-Yan Liu}.}
  \bibinfo{year}{2017}\natexlab{}.
\newblock \showarticletitle{Lightgbm: A Highly Efficient Gradient Boosting
  Decision Tree}. In \bibinfo{booktitle}{\emph{Advances in Neural Information
  Processing Systems}}. \bibinfo{pages}{3146--3154}.
\newblock


\bibitem[\protect\citeauthoryear{King, Moskowitz, Egilmez, Zhang, Zhang, Bass,
  Rogers, Ghaffari, Wakschlag, and Alshurafa}{King et~al\mbox{.}}{2019}]%
        {king2019micro}
\bibfield{author}{\bibinfo{person}{Zachary~D King}, \bibinfo{person}{Judith
  Moskowitz}, \bibinfo{person}{Begum Egilmez}, \bibinfo{person}{Shibo Zhang},
  \bibinfo{person}{Lida Zhang}, \bibinfo{person}{Michael Bass},
  \bibinfo{person}{John Rogers}, \bibinfo{person}{Roozbeh Ghaffari},
  \bibinfo{person}{Laurie Wakschlag}, {and} \bibinfo{person}{Nabil Alshurafa}.}
  \bibinfo{year}{2019}\natexlab{}.
\newblock \showarticletitle{micro-Stress EMA: A Passive Sensing Framework for
  Detecting in-the-wild Stress in Pregnant Mothers}.
\newblock \bibinfo{journal}{\emph{Proceedings of the ACM on Interactive,
  Mobile, Wearable and Ubiquitous Technologies}} \bibinfo{volume}{3},
  \bibinfo{number}{3} (\bibinfo{year}{2019}), \bibinfo{pages}{1--22}.
\newblock


\bibitem[\protect\citeauthoryear{Kj{\ae}rgaard, Ardakanian, Carlucci, Dong,
  Firth, Gao, Huebner, Mahdavi, Rahaman, Salim, et~al\mbox{.}}{Kj{\ae}rgaard
  et~al\mbox{.}}{2020}]%
        {kjaergaard2020current}
\bibfield{author}{\bibinfo{person}{Mikkel~B Kj{\ae}rgaard},
  \bibinfo{person}{Omid Ardakanian}, \bibinfo{person}{Salvatore Carlucci},
  \bibinfo{person}{Bing Dong}, \bibinfo{person}{Steven~K Firth},
  \bibinfo{person}{Nan Gao}, \bibinfo{person}{Gesche~Margarethe Huebner},
  \bibinfo{person}{Ardeshir Mahdavi}, \bibinfo{person}{Mohammad~Saiedur
  Rahaman}, \bibinfo{person}{Flora~D Salim}, {et~al\mbox{.}}}
  \bibinfo{year}{2020}\natexlab{}.
\newblock \showarticletitle{Current Practices and Infrastructure for Open Data
  based Research on Occupant-centric Design and Operation of Buildings}.
\newblock \bibinfo{journal}{\emph{Building and Environment}}
  (\bibinfo{year}{2020}), \bibinfo{pages}{106848}.
\newblock


\bibitem[\protect\citeauthoryear{Latulipe, Carroll, and Lottridge}{Latulipe
  et~al\mbox{.}}{2011}]%
        {latulipe2011love}
\bibfield{author}{\bibinfo{person}{Celine Latulipe}, \bibinfo{person}{Erin~A
  Carroll}, {and} \bibinfo{person}{Danielle Lottridge}.}
  \bibinfo{year}{2011}\natexlab{}.
\newblock \showarticletitle{Love, Hate, Arousal and Engagement: Exploring
  Audience Responses to Performing Arts}. In
  \bibinfo{booktitle}{\emph{Proceedings of the SIGCHI Conference on Human
  Factors in Computing Systems}}. ACM, \bibinfo{pages}{1845--1854}.
\newblock


\bibitem[\protect\citeauthoryear{Luque-Casado, Zabala, Morales, Mateo-March,
  and Sanabria}{Luque-Casado et~al\mbox{.}}{2013}]%
        {luque2013cognitive}
\bibfield{author}{\bibinfo{person}{Antonio Luque-Casado},
  \bibinfo{person}{Mikel Zabala}, \bibinfo{person}{Esther Morales},
  \bibinfo{person}{Manuel Mateo-March}, {and} \bibinfo{person}{Daniel
  Sanabria}.} \bibinfo{year}{2013}\natexlab{}.
\newblock \showarticletitle{Cognitive Performance and Heart Rate Variability:
  the Influence of Fitness Level}.
\newblock \bibinfo{journal}{\emph{PloS one}} \bibinfo{volume}{8},
  \bibinfo{number}{2} (\bibinfo{year}{2013}), \bibinfo{pages}{e56935}.
\newblock


\bibitem[\protect\citeauthoryear{Malhotra}{Malhotra}{2008}]%
        {malhotra2008completion}
\bibfield{author}{\bibinfo{person}{Neil Malhotra}.}
  \bibinfo{year}{2008}\natexlab{}.
\newblock \showarticletitle{Completion Time and Response Order Effects in Web
  Surveys}.
\newblock \bibinfo{journal}{\emph{Public Opinion Quarterly}}
  \bibinfo{volume}{72}, \bibinfo{number}{5} (\bibinfo{year}{2008}),
  \bibinfo{pages}{914--934}.
\newblock


\bibitem[\protect\citeauthoryear{Marks}{Marks}{2000}]%
        {marks2000student}
\bibfield{author}{\bibinfo{person}{Helen~M Marks}.}
  \bibinfo{year}{2000}\natexlab{}.
\newblock \showarticletitle{Student Engagement in Instructional Activity:
  Patterns in the Elementary, Middle, and High school years}.
\newblock \bibinfo{journal}{\emph{American educational research journal}}
  \bibinfo{volume}{37}, \bibinfo{number}{1} (\bibinfo{year}{2000}),
  \bibinfo{pages}{153--184}.
\newblock


\bibitem[\protect\citeauthoryear{Martin and Torres}{Martin and Torres}{2016}]%
        {martin2016student}
\bibfield{author}{\bibinfo{person}{Jonathan Martin} {and}
  \bibinfo{person}{Amada Torres}.} \bibinfo{year}{2016}\natexlab{}.
\newblock \showarticletitle{What is Student Engagement and Why is it
  Important}.
\newblock \bibinfo{journal}{\emph{Retrieved May}}  \bibinfo{volume}{4}
  (\bibinfo{year}{2016}), \bibinfo{pages}{2018}.
\newblock


\bibitem[\protect\citeauthoryear{McNeal, Spry, Mitra, and Tipton}{McNeal
  et~al\mbox{.}}{2014}]%
        {mcneal2014measuring}
\bibfield{author}{\bibinfo{person}{Karen~S McNeal}, \bibinfo{person}{Jacob~M
  Spry}, \bibinfo{person}{Ritayan Mitra}, {and} \bibinfo{person}{Jamie~L
  Tipton}.} \bibinfo{year}{2014}\natexlab{}.
\newblock \showarticletitle{Measuring Student Engagement, Knowledge, and
  Perceptions of Climate Change in an Introductory Environmental Geology
  Course}.
\newblock \bibinfo{journal}{\emph{Journal of Geoscience Education}}
  \bibinfo{volume}{62}, \bibinfo{number}{4} (\bibinfo{year}{2014}),
  \bibinfo{pages}{655--667}.
\newblock


\bibitem[\protect\citeauthoryear{Mehler, Reimer, and Wang}{Mehler
  et~al\mbox{.}}{2011}]%
        {mehler_reimer_wang_2011}
\bibfield{author}{\bibinfo{person}{Bruce Mehler}, \bibinfo{person}{Bryan
  Reimer}, {and} \bibinfo{person}{Ying Wang}.} \bibinfo{year}{2011}\natexlab{}.
\newblock \showarticletitle{A Comparison of Heart Rate and Heart Rate
  Variability Indices in Distinguishing Single-Task Driving and Driving Under
  Secondary Cognitive Workload}.
\newblock \bibinfo{journal}{\emph{Proceedings of the 6th International Driving
  Symposium on Human Factors in Driver Assessment, Training, and Vehicle Design
  : driving assessment 2011}} (\bibinfo{year}{2011}).
\newblock
\urldef\tempurl%
\url{https://doi.org/10.17077/drivingassessment.1451}
\showDOI{\tempurl}


\bibitem[\protect\citeauthoryear{Mendes}{Mendes}{2009}]%
        {mendes2009assessing}
\bibfield{author}{\bibinfo{person}{Wendy~Berry Mendes}.}
  \bibinfo{year}{2009}\natexlab{}.
\newblock \showarticletitle{Assessing Autonomic Nervous System Activity}.
\newblock \bibinfo{journal}{\emph{Methods in social neuroscience}}
  (\bibinfo{year}{2009}), \bibinfo{pages}{118--147}.
\newblock


\bibitem[\protect\citeauthoryear{Monkaresi, Bosch, Calvo, and
  D'Mello}{Monkaresi et~al\mbox{.}}{2016}]%
        {monkaresi2016automated}
\bibfield{author}{\bibinfo{person}{Hamed Monkaresi}, \bibinfo{person}{Nigel
  Bosch}, \bibinfo{person}{Rafael~A Calvo}, {and} \bibinfo{person}{Sidney~K
  D'Mello}.} \bibinfo{year}{2016}\natexlab{}.
\newblock \showarticletitle{Automated Detection of Engagement Using Video-based
  Estimation of Facial Expressions and Heart Rate}.
\newblock \bibinfo{journal}{\emph{IEEE Transactions on Affective Computing}}
  \bibinfo{volume}{8}, \bibinfo{number}{1} (\bibinfo{year}{2016}),
  \bibinfo{pages}{15--28}.
\newblock


\bibitem[\protect\citeauthoryear{Moore and Lippman}{Moore and Lippman}{2005}]%
        {moore2005conceptualizing}
\bibfield{author}{\bibinfo{person}{KA Moore} {and} \bibinfo{person}{L
  Lippman}.} \bibinfo{year}{2005}\natexlab{}.
\newblock \bibinfo{title}{Conceptualizing and Measuring Indicators of positive
  Development: What do Children Need to Flourish}.
\newblock
\newblock


\bibitem[\protect\citeauthoryear{Morshed, Saha, Li, D'Mello, De~Choudhury,
  Abowd, and Pl{\"o}tz}{Morshed et~al\mbox{.}}{2019}]%
        {morshed2019prediction}
\bibfield{author}{\bibinfo{person}{Mehrab~Bin Morshed},
  \bibinfo{person}{Koustuv Saha}, \bibinfo{person}{Richard Li},
  \bibinfo{person}{Sidney~K D'Mello}, \bibinfo{person}{Munmun De~Choudhury},
  \bibinfo{person}{Gregory~D Abowd}, {and} \bibinfo{person}{Thomas Pl{\"o}tz}.}
  \bibinfo{year}{2019}\natexlab{}.
\newblock \showarticletitle{Prediction of Mood Instability with Passive
  Sensing}.
\newblock \bibinfo{journal}{\emph{Proceedings of the ACM on Interactive,
  Mobile, Wearable and Ubiquitous Technologies}} \bibinfo{volume}{3},
  \bibinfo{number}{3} (\bibinfo{year}{2019}), \bibinfo{pages}{75}.
\newblock


\bibitem[\protect\citeauthoryear{M{\"u}ller, Guido, et~al\mbox{.}}{M{\"u}ller
  et~al\mbox{.}}{2016}]%
        {muller2016introduction}
\bibfield{author}{\bibinfo{person}{Andreas~C M{\"u}ller},
  \bibinfo{person}{Sarah Guido}, {et~al\mbox{.}}}
  \bibinfo{year}{2016}\natexlab{}.
\newblock \bibinfo{booktitle}{\emph{Introduction to Machine Learning with
  Python: A Guide for data scientists}}.
\newblock \bibinfo{publisher}{" O'Reilly Media, Inc."}.
\newblock


\bibitem[\protect\citeauthoryear{Nagendra, Kumar, and Mukherjee}{Nagendra
  et~al\mbox{.}}{2015}]%
        {nagendra2015cognitive}
\bibfield{author}{\bibinfo{person}{H Nagendra}, \bibinfo{person}{Vinod Kumar},
  {and} \bibinfo{person}{Shaktidev Mukherjee}.}
  \bibinfo{year}{2015}\natexlab{}.
\newblock \showarticletitle{Cognitive Behavior Evaluation Based on
  Physiological Parameters Among Young Healthy Subjects with Yoga as
  Intervention}.
\newblock \bibinfo{journal}{\emph{Computational and mathematical methods in
  medicine}}  \bibinfo{volume}{2015} (\bibinfo{year}{2015}).
\newblock


\bibitem[\protect\citeauthoryear{Nickel and Nachreiner}{Nickel and
  Nachreiner}{2003}]%
        {nickel2003sensitivity}
\bibfield{author}{\bibinfo{person}{Peter Nickel} {and}
  \bibinfo{person}{Friedhelm Nachreiner}.} \bibinfo{year}{2003}\natexlab{}.
\newblock \showarticletitle{Sensitivity and Diagnosticity of the 0.1-Hz
  component of Heart Rate Variability as an Indicator of Mental Workload}.
\newblock \bibinfo{journal}{\emph{Human factors}} \bibinfo{volume}{45},
  \bibinfo{number}{4} (\bibinfo{year}{2003}), \bibinfo{pages}{575--590}.
\newblock


\bibitem[\protect\citeauthoryear{Organization et~al\mbox{.}}{Organization
  et~al\mbox{.}}{2007}]%
        {world2007housing}
\bibfield{author}{\bibinfo{person}{World~Health Organization} {et~al\mbox{.}}}
  \bibinfo{year}{2007}\natexlab{}.
\newblock \showarticletitle{Housing, Energy, and Thermal Comfort}.
\newblock \bibinfo{journal}{\emph{A review of}}  \bibinfo{volume}{10}
  (\bibinfo{year}{2007}).
\newblock


\bibitem[\protect\citeauthoryear{Pal and Bharati}{Pal and Bharati}{2019}]%
        {pal2019applications}
\bibfield{author}{\bibinfo{person}{Manoranjan Pal} {and}
  \bibinfo{person}{Premananda Bharati}.} \bibinfo{year}{2019}\natexlab{}.
\newblock \bibinfo{booktitle}{\emph{Applications of Regression Techniques}}.
\newblock \bibinfo{publisher}{Springer}.
\newblock


\bibitem[\protect\citeauthoryear{Palumbo, Marraccini, Weyandt, Wilder-Smith,
  McGee, Liu, and Goodwin}{Palumbo et~al\mbox{.}}{2017}]%
        {palumbo2017interpersonal}
\bibfield{author}{\bibinfo{person}{Richard~V Palumbo},
  \bibinfo{person}{Marisa~E Marraccini}, \bibinfo{person}{Lisa~L Weyandt},
  \bibinfo{person}{Oliver Wilder-Smith}, \bibinfo{person}{Heather~A McGee},
  \bibinfo{person}{Siwei Liu}, {and} \bibinfo{person}{Matthew~S Goodwin}.}
  \bibinfo{year}{2017}\natexlab{}.
\newblock \showarticletitle{Interpersonal Autonomic Physiology: A Systematic
  Review of the Literature}.
\newblock \bibinfo{journal}{\emph{Personality and Social Psychology Review}}
  \bibinfo{volume}{21}, \bibinfo{number}{2} (\bibinfo{year}{2017}),
  \bibinfo{pages}{99--141}.
\newblock


\bibitem[\protect\citeauthoryear{Pawel, José, and Bahnfleth}{Pawel
  et~al\mbox{.}}{2017}]%
        {wargocki2017quantitative}
\bibfield{author}{\bibinfo{person}{Wargocki Pawel}, \bibinfo{person}{Alí
  Porras-Salazar José}, {and} \bibinfo{person}{William~P. Bahnfleth}.}
  \bibinfo{year}{2017}\natexlab{}.
\newblock \showarticletitle{Quantitative Relationships Between Classroom CO2
  Concentration and Learning in Elementary Schools}.
\newblock \bibinfo{journal}{\emph{8th AIVC Conference "Ventilating healthy
  low-energy buildings"}} (\bibinfo{year}{2017}).
\newblock


\bibitem[\protect\citeauthoryear{Pflanzer and McMullen}{Pflanzer and
  McMullen}{2013}]%
        {pflanzer2013galvanic}
\bibfield{author}{\bibinfo{person}{Richard Pflanzer} {and} \bibinfo{person}{W
  McMullen}.} \bibinfo{year}{2013}\natexlab{}.
\newblock \showarticletitle{Galvanic Skin Response and the Polygraph}.
\newblock \bibinfo{journal}{\emph{BIOPAC Systems, Inc. Retrieved}}
  \bibinfo{volume}{5} (\bibinfo{year}{2013}).
\newblock


\bibitem[\protect\citeauthoryear{Pintrich and De~Groot}{Pintrich and
  De~Groot}{1990}]%
        {pintrich1990motivational}
\bibfield{author}{\bibinfo{person}{Paul~R Pintrich} {and}
  \bibinfo{person}{Elisabeth~V De~Groot}.} \bibinfo{year}{1990}\natexlab{}.
\newblock \showarticletitle{Motivational and Self-regulated Learning Components
  of Classroom Academic Performance.}
\newblock \bibinfo{journal}{\emph{Journal of educational psychology}}
  \bibinfo{volume}{82}, \bibinfo{number}{1} (\bibinfo{year}{1990}),
  \bibinfo{pages}{33}.
\newblock


\bibitem[\protect\citeauthoryear{Pollak, Adams, and Gay}{Pollak
  et~al\mbox{.}}{2011}]%
        {pollak2011pam}
\bibfield{author}{\bibinfo{person}{John~P Pollak}, \bibinfo{person}{Phil
  Adams}, {and} \bibinfo{person}{Geri Gay}.} \bibinfo{year}{2011}\natexlab{}.
\newblock \showarticletitle{PAM: A Photographic Affect Meter for Frequent, in
  Situ Measurement of Affect}. In \bibinfo{booktitle}{\emph{Proceedings of the
  SIGCHI conference on Human factors in computing systems}}. ACM,
  \bibinfo{pages}{725--734}.
\newblock


\bibitem[\protect\citeauthoryear{Rahaman, Liono, Ren, Chan, Kudo, Rawling, and
  Salim}{Rahaman et~al\mbox{.}}{2020}]%
        {rahaman2020ambient}
\bibfield{author}{\bibinfo{person}{Mohammad~Saiedur Rahaman},
  \bibinfo{person}{Jonathan Liono}, \bibinfo{person}{Yongli Ren},
  \bibinfo{person}{Jeffrey Chan}, \bibinfo{person}{Shaw Kudo},
  \bibinfo{person}{Tim Rawling}, {and} \bibinfo{person}{Flora~D Salim}.}
  \bibinfo{year}{2020}\natexlab{}.
\newblock \showarticletitle{An Ambient-Physical System to Infer Concentration
  in Open-plan Workplace}.
\newblock \bibinfo{journal}{\emph{IEEE Internet of Things Journal}}
  (\bibinfo{year}{2020}).
\newblock


\bibitem[\protect\citeauthoryear{Rivera-Pelayo, Zacharias, M{\"u}ller, and
  Braun}{Rivera-Pelayo et~al\mbox{.}}{2012}]%
        {rivera2012applying}
\bibfield{author}{\bibinfo{person}{Ver{\'o}nica Rivera-Pelayo},
  \bibinfo{person}{Valentin Zacharias}, \bibinfo{person}{Lars M{\"u}ller},
  {and} \bibinfo{person}{Simone Braun}.} \bibinfo{year}{2012}\natexlab{}.
\newblock \showarticletitle{Applying Quantified Self Approaches to Support
  Reflective Learning}. In \bibinfo{booktitle}{\emph{Proceedings of the 2nd
  international conference on learning analytics and knowledge}}.
  \bibinfo{pages}{111--114}.
\newblock


\bibitem[\protect\citeauthoryear{Sadri, Ren, and Salim}{Sadri
  et~al\mbox{.}}{2017}]%
        {sadri2017information}
\bibfield{author}{\bibinfo{person}{Amin Sadri}, \bibinfo{person}{Yongli Ren},
  {and} \bibinfo{person}{Flora~D Salim}.} \bibinfo{year}{2017}\natexlab{}.
\newblock \showarticletitle{Information Gain-based Metric for Recognizing
  Transitions in Human Activities}.
\newblock \bibinfo{journal}{\emph{Pervasive and Mobile Computing}}
  \bibinfo{volume}{38} (\bibinfo{year}{2017}), \bibinfo{pages}{92--109}.
\newblock


\bibitem[\protect\citeauthoryear{Sarchiapone, Gramaglia, Iosue, Carli,
  Mandelli, Serretti, Marangon, and Zeppegno}{Sarchiapone
  et~al\mbox{.}}{2018}]%
        {sarchiapone2018association}
\bibfield{author}{\bibinfo{person}{Marco Sarchiapone}, \bibinfo{person}{Carla
  Gramaglia}, \bibinfo{person}{Miriam Iosue}, \bibinfo{person}{Vladimir Carli},
  \bibinfo{person}{Laura Mandelli}, \bibinfo{person}{Alessandro Serretti},
  \bibinfo{person}{Debora Marangon}, {and} \bibinfo{person}{Patrizia
  Zeppegno}.} \bibinfo{year}{2018}\natexlab{}.
\newblock \showarticletitle{The Association Between Electrodermal Activity
  (EDA), Depression and Suicidal Behaviour: A Systematic Review and Narrative
  Synthesis}.
\newblock \bibinfo{journal}{\emph{BMC psychiatry}} \bibinfo{volume}{18},
  \bibinfo{number}{1} (\bibinfo{year}{2018}), \bibinfo{pages}{22}.
\newblock


\bibitem[\protect\citeauthoryear{Satish, Mendell, Shekhar, Hotchi, Sullivan,
  Streufert, and Fisk}{Satish et~al\mbox{.}}{2012}]%
        {satish2012co2}
\bibfield{author}{\bibinfo{person}{Usha Satish}, \bibinfo{person}{Mark~J
  Mendell}, \bibinfo{person}{Krishnamurthy Shekhar}, \bibinfo{person}{Toshifumi
  Hotchi}, \bibinfo{person}{Douglas Sullivan}, \bibinfo{person}{Siegfried
  Streufert}, {and} \bibinfo{person}{William~J Fisk}.}
  \bibinfo{year}{2012}\natexlab{}.
\newblock \showarticletitle{Is CO2 an Indoor Pollutant? Direct Effects of
  Low-to-moderate CO2 Cncentrations on Human Decision-making Performance}.
\newblock \bibinfo{journal}{\emph{Environmental health perspectives}}
  \bibinfo{volume}{120}, \bibinfo{number}{12} (\bibinfo{year}{2012}),
  \bibinfo{pages}{1671--1677}.
\newblock


\bibitem[\protect\citeauthoryear{Seber and Lee}{Seber and Lee}{2012}]%
        {seber2012linear}
\bibfield{author}{\bibinfo{person}{George~AF Seber} {and}
  \bibinfo{person}{Alan~J Lee}.} \bibinfo{year}{2012}\natexlab{}.
\newblock \bibinfo{booktitle}{\emph{Linear Regression Analysis}}.
  Vol.~\bibinfo{volume}{329}.
\newblock \bibinfo{publisher}{John Wiley \& Sons}.
\newblock


\bibitem[\protect\citeauthoryear{Senin}{Senin}{2008}]%
        {senin2008dynamic}
\bibfield{author}{\bibinfo{person}{Pavel Senin}.}
  \bibinfo{year}{2008}\natexlab{}.
\newblock \showarticletitle{Dynamic Time Warping Algorithm Review}.
\newblock \bibinfo{journal}{\emph{Information and Computer Science Department
  University of Hawaii at Manoa Honolulu, USA}} \bibinfo{volume}{855},
  \bibinfo{number}{1-23} (\bibinfo{year}{2008}), \bibinfo{pages}{40}.
\newblock


\bibitem[\protect\citeauthoryear{Shaffer and Ginsberg}{Shaffer and
  Ginsberg}{2017}]%
        {shaffer2017overview}
\bibfield{author}{\bibinfo{person}{Fred Shaffer} {and} \bibinfo{person}{JP
  Ginsberg}.} \bibinfo{year}{2017}\natexlab{}.
\newblock \showarticletitle{An Overview of Heart Rate Variability Metrics and
  Norms}.
\newblock \bibinfo{journal}{\emph{Frontiers in public health}}
  \bibinfo{volume}{5} (\bibinfo{year}{2017}), \bibinfo{pages}{258}.
\newblock


\bibitem[\protect\citeauthoryear{Shao, Prabowo, Zhao, Tan, Koniusz, Chan, Hei,
  Feest, and Salim}{Shao et~al\mbox{.}}{2019}]%
        {shao2019flight}
\bibfield{author}{\bibinfo{person}{Wei Shao}, \bibinfo{person}{Arian Prabowo},
  \bibinfo{person}{Sichen Zhao}, \bibinfo{person}{Siyu Tan},
  \bibinfo{person}{Piotr Koniusz}, \bibinfo{person}{Jeffrey Chan},
  \bibinfo{person}{Xinhong Hei}, \bibinfo{person}{Bradley Feest}, {and}
  \bibinfo{person}{Flora~D Salim}.} \bibinfo{year}{2019}\natexlab{}.
\newblock \showarticletitle{Flight Delay Prediction using Airport Situational
  Awareness Map}. In \bibinfo{booktitle}{\emph{Proceedings of the 27th ACM
  SIGSPATIAL International Conference on Advances in Geographic Information
  Systems}}. \bibinfo{pages}{432--435}.
\newblock


\bibitem[\protect\citeauthoryear{Shernoff, Csikszentmihalyi, Schneider, and
  Shernoff}{Shernoff et~al\mbox{.}}{2014}]%
        {shernoff2014student}
\bibfield{author}{\bibinfo{person}{David~J Shernoff}, \bibinfo{person}{Mihaly
  Csikszentmihalyi}, \bibinfo{person}{Barbara Schneider}, {and}
  \bibinfo{person}{Elisa~Steele Shernoff}.} \bibinfo{year}{2014}\natexlab{}.
\newblock \showarticletitle{Student Engagement in High School Classrooms from
  the Perspective of Flow Theory}.
\newblock In \bibinfo{booktitle}{\emph{Applications of flow in human
  development and education}}. \bibinfo{publisher}{Springer},
  \bibinfo{pages}{475--494}.
\newblock


\bibitem[\protect\citeauthoryear{Skinner, Furrer, Marchand, and
  Kindermann}{Skinner et~al\mbox{.}}{2008}]%
        {skinner2008engagement}
\bibfield{author}{\bibinfo{person}{Ellen Skinner}, \bibinfo{person}{Carrie
  Furrer}, \bibinfo{person}{Gwen Marchand}, {and} \bibinfo{person}{Thomas
  Kindermann}.} \bibinfo{year}{2008}\natexlab{}.
\newblock \showarticletitle{Engagement and Disaffection in the Classroom: Part
  of a Larger Motivational Dynamic?}
\newblock \bibinfo{journal}{\emph{Journal of educational psychology}}
  \bibinfo{volume}{100}, \bibinfo{number}{4} (\bibinfo{year}{2008}),
  \bibinfo{pages}{765}.
\newblock


\bibitem[\protect\citeauthoryear{Skinner, Kindermann, and Furrer}{Skinner
  et~al\mbox{.}}{2009}]%
        {skinner2009motivational}
\bibfield{author}{\bibinfo{person}{Ellen~A Skinner}, \bibinfo{person}{Thomas~A
  Kindermann}, {and} \bibinfo{person}{Carrie~J Furrer}.}
  \bibinfo{year}{2009}\natexlab{}.
\newblock \showarticletitle{A Motivational Perspective on Engagement and
  Disaffection: Conceptualization and Assessment of Children's Behavioral and
  Emotional Participation in Academic Activities in the Classroom}.
\newblock \bibinfo{journal}{\emph{Educational and Psychological Measurement}}
  \bibinfo{volume}{69}, \bibinfo{number}{3} (\bibinfo{year}{2009}),
  \bibinfo{pages}{493--525}.
\newblock


\bibitem[\protect\citeauthoryear{Stipek}{Stipek}{2002}]%
        {stipek2002good}
\bibfield{author}{\bibinfo{person}{Deborah Stipek}.}
  \bibinfo{year}{2002}\natexlab{}.
\newblock \showarticletitle{Good Instruction is Motivating}.
\newblock In \bibinfo{booktitle}{\emph{Development of achievement motivation}}.
  \bibinfo{publisher}{Elsevier}, \bibinfo{pages}{309--332}.
\newblock


\bibitem[\protect\citeauthoryear{Tanner}{Tanner}{2013}]%
        {tanner2013structure}
\bibfield{author}{\bibinfo{person}{Kimberly~D Tanner}.}
  \bibinfo{year}{2013}\natexlab{}.
\newblock \showarticletitle{Structure Matters: Twenty-one Teaching Strategies
  to Promote Student Engagement and Cultivate Classroom Equity}.
\newblock \bibinfo{journal}{\emph{CBE—Life Sciences Education}}
  \bibinfo{volume}{12}, \bibinfo{number}{3} (\bibinfo{year}{2013}),
  \bibinfo{pages}{322--331}.
\newblock


\bibitem[\protect\citeauthoryear{van Gent, Farah, van Nes, and van Arem}{van
  Gent et~al\mbox{.}}{2019}]%
        {van2019heartpy}
\bibfield{author}{\bibinfo{person}{Paul van Gent}, \bibinfo{person}{Haneen
  Farah}, \bibinfo{person}{Nicole van Nes}, {and} \bibinfo{person}{Bart van
  Arem}.} \bibinfo{year}{2019}\natexlab{}.
\newblock \showarticletitle{HeartPy: A Novel Heart Rate Algorithm for the
  Analysis of Noisy Signals}.
\newblock \bibinfo{journal}{\emph{Transportation research part F: traffic
  psychology and behaviour}}  \bibinfo{volume}{66} (\bibinfo{year}{2019}),
  \bibinfo{pages}{368--378}.
\newblock


\bibitem[\protect\citeauthoryear{von Zimmermann, Vicary, Sperling, Orgs, and
  Richardson}{von Zimmermann et~al\mbox{.}}{2018}]%
        {von2018choreography}
\bibfield{author}{\bibinfo{person}{Jorina von Zimmermann},
  \bibinfo{person}{Staci Vicary}, \bibinfo{person}{Matthias Sperling},
  \bibinfo{person}{Guido Orgs}, {and} \bibinfo{person}{Daniel~C Richardson}.}
  \bibinfo{year}{2018}\natexlab{}.
\newblock \showarticletitle{The Choreography of Group Affiliation}.
\newblock \bibinfo{journal}{\emph{Topics in Cognitive Science}}
  \bibinfo{volume}{10}, \bibinfo{number}{1} (\bibinfo{year}{2018}),
  \bibinfo{pages}{80--94}.
\newblock


\bibitem[\protect\citeauthoryear{Wang and Cesar}{Wang and Cesar}{2015}]%
        {wang2015physiological}
\bibfield{author}{\bibinfo{person}{Chen Wang} {and} \bibinfo{person}{Pablo
  Cesar}.} \bibinfo{year}{2015}\natexlab{}.
\newblock \showarticletitle{Physiological Measurement on Students' Engagement
  in a Distributed Learning Environment}. In \bibinfo{booktitle}{\emph{PhyCS}}.
  \bibinfo{pages}{149--156}.
\newblock


\bibitem[\protect\citeauthoryear{Wang, Chen, Chen, Li, Harari, Tignor, Zhou,
  Ben-Zeev, and Campbell}{Wang et~al\mbox{.}}{2014}]%
        {wang2014studentlife}
\bibfield{author}{\bibinfo{person}{Rui Wang}, \bibinfo{person}{Fanglin Chen},
  \bibinfo{person}{Zhenyu Chen}, \bibinfo{person}{Tianxing Li},
  \bibinfo{person}{Gabriella Harari}, \bibinfo{person}{Stefanie Tignor},
  \bibinfo{person}{Xia Zhou}, \bibinfo{person}{Dror Ben-Zeev}, {and}
  \bibinfo{person}{Andrew~T Campbell}.} \bibinfo{year}{2014}\natexlab{}.
\newblock \showarticletitle{StudentLife: Assessing Mental Health, Academic
  Performance and Behavioral Trends of College Students Using Smartphones}. In
  \bibinfo{booktitle}{\emph{Proceedings of the 2014 ACM international joint
  conference on pervasive and ubiquitous computing}}. ACM,
  \bibinfo{pages}{3--14}.
\newblock


\bibitem[\protect\citeauthoryear{Wang, Harari, Wang, M{\"u}ller, Mirjafari,
  Masaba, and Campbell}{Wang et~al\mbox{.}}{2018}]%
        {wang2018sensing}
\bibfield{author}{\bibinfo{person}{Weichen Wang}, \bibinfo{person}{Gabriella~M
  Harari}, \bibinfo{person}{Rui Wang}, \bibinfo{person}{Sandrine~R M{\"u}ller},
  \bibinfo{person}{Shayan Mirjafari}, \bibinfo{person}{Kizito Masaba}, {and}
  \bibinfo{person}{Andrew~T Campbell}.} \bibinfo{year}{2018}\natexlab{}.
\newblock \showarticletitle{Sensing Behavioral Change over Time: Using
  Within-person Variability Features from Mobile Sensing to Predict Personality
  Traits}.
\newblock \bibinfo{journal}{\emph{Proceedings of the ACM on Interactive,
  Mobile, Wearable and Ubiquitous Technologies}} \bibinfo{volume}{2},
  \bibinfo{number}{3} (\bibinfo{year}{2018}), \bibinfo{pages}{141}.
\newblock


\bibitem[\protect\citeauthoryear{Ward, Richardson, Orgs, Hunter, and
  Hamilton}{Ward et~al\mbox{.}}{2018}]%
        {ward2018sensing}
\bibfield{author}{\bibinfo{person}{Jamie~A Ward}, \bibinfo{person}{Daniel
  Richardson}, \bibinfo{person}{Guido Orgs}, \bibinfo{person}{Kelly Hunter},
  {and} \bibinfo{person}{Antonia Hamilton}.} \bibinfo{year}{2018}\natexlab{}.
\newblock \showarticletitle{Sensing Interpersonal Synchrony Between Actors and
  Autistic Children in Theatre Using Wrist-worn Accelerometers}. In
  \bibinfo{booktitle}{\emph{Proceedings of the 2018 ACM International Symposium
  on Wearable Computers}}. ACM, \bibinfo{pages}{148--155}.
\newblock


\bibitem[\protect\citeauthoryear{Wiens, Dale, Boyce, and Kershaw}{Wiens
  et~al\mbox{.}}{2008}]%
        {wiens2008three}
\bibfield{author}{\bibinfo{person}{Trevor~S Wiens}, \bibinfo{person}{Brenda~C
  Dale}, \bibinfo{person}{Mark~S Boyce}, {and} \bibinfo{person}{G~Peter
  Kershaw}.} \bibinfo{year}{2008}\natexlab{}.
\newblock \showarticletitle{Three Way K-fold Cross-validation of Resource
  Selection Functions}.
\newblock \bibinfo{journal}{\emph{Ecological Modelling}} \bibinfo{volume}{212},
  \bibinfo{number}{3-4} (\bibinfo{year}{2008}), \bibinfo{pages}{244--255}.
\newblock


\end{thebibliography}

\end{document}